\documentclass[12pt,a4paper]{article}
\usepackage{amssymb}
\newcommand{\Nis}{n_c}
\newcommand{\acom}[1]{\{\,#1\,\}}
\newcommand{\R}{\mathbb{R}}
\newcommand{\Z}{\mathbb{Z}}
\newcommand{\N}{\mathbb{N}}
\newcommand{\epn}{\epsilon}
\newcommand{\p}{\partial}
\newcommand{\T}[1]{{\mathring{#1}}}
\newcommand{\nn}{\nonumber\\}
\newcommand{\gIIB}{g_s^{\rm{IIB}}}
\newcommand{\gsb}{g_b}
\newcommand{\cpq}{c_{pq}}
\newcommand{\Opq}{O_{(p,q)}}
\newcommand{\hatg}{\hat{\gamma}}
\newcommand{\dkp}{\delta_\kappa}
\newcommand{\dss}{\delta_\epsilon}

\newcommand{\hi}{\hat{i}}
\newcommand{\hj}{\hat{j}}
\newcommand{\hk}{\hat{k}}
\newcommand{\hM}{\hat{M}}
\newcommand{\hN}{\hat{N}}
\newcommand{\hP}{\hat{P}}
\newcommand{\hA}{\hat{A}}
\newcommand{\tA}{\tilde{A}}
\newcommand{\thA}{\skew{5}\tilde{\hat{A}}}
\newcommand{\hF}{\hat{F}}
\newcommand{\tF}{\tilde{F}}
\newcommand{\thF}{\skew{3}\tilde{\hat{F}}}
\newcommand{\tH}{\tilde{H}}
\newcommand{\hB}{\hat{B}}
\newcommand{\hC}{\hat{C}}
\newcommand{\RM}{{\tilde{M}}}
\newcommand{\RN}{{\tilde{N}}}
\newcommand{\RP}{{\tilde{P}}}
\newcommand{\RQ}{{\tilde{Q}}}
\newcommand{\hmu}{\hat{\mu}}
\newcommand{\hnu}{\hat{\nu}}
\newcommand{\hrho}{\hat{\rho}}
\newcommand{\hsigma}{\hat{\sigma}}
\newcommand{\htau}{\hat{\tau}}
\newcommand{\hr}{\hat{r}}
\newcommand{\hs}{\hat{s}}
\newcommand{\hrstu}{\hat{r}\hat{s}\hat{t}\hat{u}}
\newcommand{\hrst}{\hat{r}\hat{s}\hat{t}}
\newcommand{\hatt}{\hat{t}}
\newcommand{\hu}{\hat{u}}
\newcommand{\hv}{\hat{v}}

\newcommand{\rmu}{\tilde{\mu}}
\newcommand{\rnu}{\tilde{\nu}}
\newcommand{\rrho}{\tilde{\rho}}
\newcommand{\rsigma}{\tilde{\sigma}}
\newcommand{\tphi}{\tilde{\phi}}
\newcommand{\GO}{\mathring{G}}

\newcommand{\hPi}{\hat{\Pi}}
\newcommand{\hE}{\hat{E}}
\newcommand{\tE}{\tilde{E}}
\newcommand{\thE}{\skew{3}\tilde{\hat{E}}}
\newcommand{\hp}{\hat{p}}
\newcommand{\hq}{\hat{q}}
\newcommand{\hG}{G}

\newcommand{\ga}{g}
\newcommand{\tg}{\tilde{g}}
\newcommand{\tG}{\tilde{G}}

\newcommand{\BNS}{B^{(1)}}
\newcommand{\BR}{B^{(2)}}
\newcommand{\Bpq}{B^{(pq)}}
\newcommand{\bpq}{b^{(\hp\hq)}}
\newcommand{\bqp}{b^{(-\hq\hp)}}
\newcommand{\thG}{\tilde{\hat{\Gamma}}}
\newcommand{\hatG}{\hat{\Gamma}}
\newcommand{\thO}{\tilde{\hat{\Omega}}}
\newcommand{\tho}{\tilde{\hat{\omega}}}
\newcommand{\tOmg}{\tilde{\Omega}}
\newcommand{\tomg}{\tilde{\omega}}

\newcommand{\tcalB}{\tilde{\mathcal{B}}}
\newcommand{\tcalG}{\tilde{\mathcal{G}}}
\newcommand{\calB}{\mathcal{B}}
\newcommand{\calG}{\mathcal{G}}

\newcommand{\tQ}{\tilde{Q}}
\newcommand{\tP}{\tilde{P}}

\newcommand{\btheta}{\theta}
\newcommand{\bbtheta}{\bar{\theta}}
\newcommand{\bkappa}{\kappa}
\newcommand{\bepn}{\epsilon}
\newcommand{\bbepn}{\bar{\epsilon}}

\newcommand{\thse}{\skew{2}\tilde{\hat{e}}}
\newcommand{\tse}{\tilde{e}}

\newcommand{\Dpq}{\Delta_{(\hp\hq)}}

\newcommand{\tm}{\tilde{m}}

\newcommand{\zl}{C_{pq}}

\newcommand{\tbG}{\tilde{\mathbb{G}}}
\newcommand{\tbB}{\tilde{\mathbb{B}}}

\newcommand{\hbC}{\hat{\mathbb{C}}}
\newcommand{\XB}{\underline{X}}
\newcommand{\DB}{\underline{D}}
\newcommand{\PB}{\underline{P}_{(+)}}
\newcommand{\nB}{\underline{\nabla}}
\newcommand{\GB}{\underline{\Gamma}}
\newcommand{\tgamma}{\tilde{\gamma}}
\newcommand{\tGamma}{\tilde{\Gamma}}
\newcommand{\tl}{\tilde{l}}

\newcommand{\ffv}{f^{(5)}}
\newcommand{\fth}{f^{(3)}}
\newcommand{\fo}{f^{(1)}}
\newcommand{\hth}{h^{(pq)}}
\newcommand{\bomg}{\omega}

\newcommand{\ydg}{g'}
\newcommand{\ydA}{A'}
\newcommand{\ydE}{E'}
\newcommand{\ydB}{B'}
\newcommand{\ydP}{P'}
\newcommand{\ydQ}{Q'}
\newcommand{\ydcalG}{\mathcal{G}'}
\newcommand{\ydcalB}{\mathcal{B}'}
\newcommand{\ydbG}{\mathbb{G}'}
\newcommand{\ydbB}{\mathbb{B}'}

\newcommand{\Op}{\Omega_{\chi}}

\newcommand{\vpc}{{\varphi_0}}
\newcommand{\pc}{{\phi_0}}
\newcommand{\lc}{{l_0}}

\newcommand{\calJ}{\mathcal{J}}
\newcommand{\pbp}{\underline{P}_+}
\newcommand{\pbm}{\underline{P}_-}
\newcommand{\pbpm}{\underline{P}_{\pm}}
\newcommand{\pbmp}{\underline{P}_{\mp}}

\title{Green-Schwarz superstring action for ($p,q$)-strings \\
	from a wrapped supermembrane on a 2-torus}
\author{
{\sc Hiroyuki Okagawa$^1$}\footnote{email:
	okagawa@eken.phys.nagoya-u.ac.jp}
 \footnote{Present address: Toshiba Corporation,Yokkaichi, 512-8550,
	Japan},~
{\sc Shozo Uehara$^2$}\footnote{e-mail:
	uehara@is.utsunomiya-u.ac.jp}~ and
{\sc Satoshi Yamada$^1$}\footnote{e-mail:
	yamada@eken.phys.nagoya-u.ac.jp}\vspace{4mm}\\
{\sl $^1$Department of Physics, Nagoya University,}\\
{\sl Chikusa-ku, Nagoya 464-8602, Japan,}\\
{\sl $^2$Department of Information Systems Science, Utsunomiya
	University,}\\
{\sl Utsunomiya 321-8585, Japan,}}
\date{}
\makeatletter

 \@addtoreset{equation}{section}
\renewcommand{\thefigure}{\@arabic\c@figure}
\makeatother
\begin{document}
\maketitle
\vspace{-85mm}
\begin{flushright}
	arXiv:0811.4657
\end{flushright}
\vspace{60mm}

\begin{abstract}
We consider a wrapped supermembrane around non-trivial two cycles of
a 2-torus.
We examine the double dimensional reduction and the T-dual
transformation to deduce Green-Schwarz type IIB superstring action for
$(p,q)$-strings directly from the wrapped supermembrane on the
2-torus.
The resulting action has the couplings with both the NSNS- and the
RR-background fields and has the tension of the $(p,q)$-string.
\end{abstract}

\newpage
\section{Introduction}
The supermembrane in eleven dimensions \cite{BST}
is expected to play an important role to understand the
fundamental degrees of freedom in M-theory.
In fact, it was shown that the wrapped supermembrane in a
$S^1$-compactified eleven dimensions is related to the type IIA
superstring in ten dimensions
by means of the double dimensional reduction \cite{DHIS}.
Meanwhile type IIB superstring is related to type IIA superstring via
T-duality, or the type IIA superstring on $\R^9\times S^1$
leads to the type IIB superstring on $\R^{10}$ in the shrinking limit
of the $S^1$.
Accordingly, the supermembrane wrapped on a vanishing 2-torus is
reduced to the type IIB superstring in ten dimensions.

The transformation rule for the NSNS fields under T-duality in type II
superstring theory is given in the sigma-model with (at least) one
isometry direction, which is the well-known Buscher's rule
\cite{Bus}.
The generalized Buscher's rule, which is the transformation rule
under T-duality not only for the NSNS fields but also for the RR
fields in type II superstring, was derived at the level of the low
energy effective action of type II string theory \cite{BHO,MO}.
In addition, the generalized Buscher's rule was also derived for
the type II Green-Schwarz superstring action \cite{CLPS}
which is obtained by means of the double dimensional reduction of
the wrapped supermembrane up to quadratic order in the
anti-commuting superspace coordinates \cite{dWPP}.

It is well-known that at low-energy level the type IIB superstring
theory has a duality group of $SL(2,\R)$ which is broken down to
$SL(2,\Z)$ by the quantum effect.
Schwarz showed an $SL(2,\Z)$ family of string solutions in type IIB
supergravity \cite{Sch}, which couples to both the NSNS and the RR
background fields.
The $(p,q)$-string \cite{Sch,W} is considered to be the bound state of
fundamental strings (F-strings) and D1-branes (D-strings) in type IIB
superstring theory.
Then, $SL(2,\Z)$-covariant string actions were proposed \cite{T,CT}.
Meanwhile, the supermembrane which is wrapping $p$-times around one of
the two compact directions and $q$-times around the other direction
gives a $(p,q)$-string, however the direct derivation of the action
was not given.
Recently the bosonic sector of the type IIB Green-Schwarz superstring
action for $(p,q)$-string was derived directly from the wrapped
supermembrane action on a 2-torus \cite{OUY1}.

In this paper we shall proceed to the analysis with the anti-commuting
superspace coordinates being recovered. We use the normal coordinates
in the superfield formulation \cite{McA,AD} of the supermembrane in a
supergravity background.
We consider a supermembrane wrapped around non-trivial two cycles on
a 2-torus.
In fact, the action is expanded with respect to the anti-commuting
coordinates $\theta$ up to quadratic order \cite{dWPP}.
We shall take the shrinking volume limit of the 2-torus to approach
type IIB superstring theory along the line of Ref.\cite{CLPS}.
Then, we deduce $(p,q)$-strings in type IIB superstring theory from
the wrapped supermembrane in the limit.
We shall see that the string carries $p$-times the unit NSNS 2-form
charge and $q$-times the unit RR 2-form charge as well, which
indicates that the deduced string is, in fact, a $(p,q)$-string in
type IIB superstring theory.

The plan of this paper is as follows.
In the next section, we set up the supermembrane compactified on $T^2$
up to quadratic order in $\theta$.
In section \ref{S:DDR} we shall carefully rewrite the
eleven-dimensional supergravity background fields compactified on a
2-torus and consider the double dimensional reduction along an oblique
direction of the 2-torus.
In section \ref{S:C}, we consider the T-dual of the derived
superstring action along the other compact direction of the 2-torus to
deduce the action of $(p,q)$-strings.
The final section contains some discussion.

\section{Supermembrane in 11-dimensional superspace}\label{S:F}
The action of a supermembrane coupled to an eleven-dimensional
supergravity background is given by \cite{BST}
\begin{eqnarray}
  S&=&T\int\!d\sigma^0 \!\int_0^{2\pi}\!d\sigma^1 d\sigma^2
   \Biggl[-\frac{1}{2}\,\sqrt{-\hatg}\,\hatg^{\hi\hj}\,
	\hPi_{\hi}^{~A} \hPi_{\hj}^{~B}\,\eta_{AB}\nn
  &&\hspace{20ex}{}+\frac{1}{2}\,\sqrt{-\hatg}
    -\frac{1}{3!}\,\epn^{\hi\hj\hk}\,\p_{\hi}Z^{\hM}\p_{\hj}
    Z^{\hN}\p_{\hk} Z^{\hP}\hC_{\hP\hN\hM}\Biggr]\label{eq:SMac},
\end{eqnarray}
where $T$ is the tension of supermembrane,\footnote{The
eleven-dimensional Planck length $l_{11}$ is defined by
$T=(2\pi)^{-2}l_{11}^{-3}$.} $\hC_{\hM\hN\hP}(Z)$ is the super
three-form,
\begin{equation}
  \hPi_{\hi}^{~\hA}=(\p_{\hi} Z^{\hM})\,\hE_{\hM}^{~\hA}\,,
\end{equation}
$\hatg_{\hi\hj}\ (\hi,\hj=0,1,2)$ is the
worldvolume metric, $\hatg=\det\hatg_{\hi\hj}$, the target space
is a supermanifold with the superspace coordinates
$Z^{\hM}=(X^M,\theta^{\alpha})~(M=0,\cdots,10,~\alpha=1,\cdots,32)$.
Furthermore, with the tangent superspace index $\hA=(A,a)$,
$\hE_{\hM}^{~\hA}$ is the supervielbein and $\eta_{AB}$ is the
tangent space metric in eleven dimensions.
The mass dimensions of the worldvolume parameters $\sigma^{\hi}$ and
the eleven-dimensional background fields ($G_{MN}$, $\hC_{MNP}$) are
$0$, while that of the worldvolume metric $\hatg_{\hi\hj}$ is $-2$.
Note that the variation w.r.t.\ $\hatg_{\hi\hj}$ yields the induced
metric
\begin{equation}
  \hatg_{\hi\hj}=\hPi_{\hi}^{~A} \hPi_{\hj}^{~B}\,\eta_{AB}\,,
  \label{eq:indm3}
\end{equation}
and plugging it back into the original action leads to the Nambu-Goto
form
\begin{equation}
  S=T\int\!d\sigma^0\!\int_0^{2\pi}\!d\sigma^1 d\sigma^2
   \Biggl[-\sqrt{-\det\,(\hPi_{\hi}^{~A}\hPi_{\hj}^{~B}\,\eta_{AB})}
    -\frac{1}{3!}\,\epn^{\hi\hj\hk}\,\p_{\hi}Z^{\hM}\p_{\hj}Z^{\hN}
    \p_{\hk}Z^{\hP}\hC_{\hP\hN\hM}\Biggr]\,.\label{eq:Mac1}
\end{equation}
In fact, it is convenient to work in the Nambu-Goto action when we
carry out the double dimensional reduction in the next section.
Note that the action (\ref{eq:Mac1}) has a fermionic gauge symmetry,
or the $\kappa$-symmetry
\begin{equation}
 \dkp Z^{\hM}\hE_{\hM}^{\,~A}=0\,,\quad
 \dkp Z^{\hM}\hE_{\hM}^{\,~a}=(1+\hatG_K)^a_{~b}\,\kappa^b\,,
 \label{eq:K11}
\end{equation}
where
\begin{equation}
  \hatG_K=\frac{1}{6}\,\frac{1}{\sqrt{-\det\hatg_{\hi\hj}}}\,
	\epn^{\hi\hj\hk}\hPi_{\hi}^{~\hA}\hPi_{\hj}^{~\hB}
	\hPi_{\hk}^{~\hC}\hatG_{\hA\hB\hC}\,,
\end{equation}
$\hatg_{\hi\hj}$ is the induced metric (\ref{eq:indm3}) and the
parameter $\kappa$ is a 32-component spacetime Majorana spinor and a
worldvolume scalar. In fact, the $\kappa$-symmetry exists provided
that the background supergeometry is constrained \cite{BST}, which is
equivalent to the on-shell D=11 supergravity background \cite{CFBH}

The supervielbein and the super three-form are explicitly given to
$O(\theta^2)$ in the fermionic coordinates \cite{dWPP}. Setting the
fermionic background fields to zero, we have\footnote{$M_\alpha^{~a}$
is of $O(\theta^2)$, however, the explicit form is not used in our
analysis.}
\begin{equation}
 \hE_M^{~A}=\hat{e}_M^{~A}
	+i\bar{\theta}{\Gamma}^A\hat{\Omega}_M\theta\,,\quad
 \hE_M^{~a}=(\hat{\Omega}_M \theta)^a\,,\quad
 \hE_\alpha^{~A}=-i(\bar{\theta}{\Gamma}^A)_\alpha\,,\quad
 \hE_\alpha^{~a}=\delta_\alpha ^a+M_\alpha^{~a}\,,\label{eq:11dhE}
\end{equation}
and\footnote{Symmetrization $[*\cdots *]$ and anti-symmetrization
$(*\cdots*)$ of the indices are made with unit weight,
$A_{[M}B_{N]}=(1/2)(A_MB_N-A_NB_M),$ etc. (see Appendix \ref{S:N}).}
\begin{eqnarray}
 &&\hC_{MNP}=\hA_{MNP}
     +3i\bar{\theta}\hatG_{[MN}\hat{\Omega}_{P]}\theta\,,
  \qquad\hC_{MN\alpha}=-i(\bar{\theta}\hatG_{MN})_\alpha\,,\nn
 &&\hC_{M\alpha\beta}=(\bar{\theta}\hatG_{MN})_{(\alpha}
 	(\bar{\theta}\hatG^N)_{\beta)}\,,
  \qquad\hC_{\alpha\beta\gamma}=i(\bar{\theta}\hatG_{MN})_{(\alpha}
	(\bar{\theta}\hatG^{M})_\beta
	(\bar{\theta}\hatG^{N})_{\gamma)}\,,\label{eq:11dhC}
\end{eqnarray}
where
\begin{eqnarray}
 \hat{\Omega}_M&=&\frac{1}{4}\,\hat{\omega}_M^{\,~BC}\Gamma_{BC}
    -\hat{T}_M\,, \label{eq:11dhO}\\
 \hat{T}_M{}&=&\frac{1}{288}\,(\hatG_M^{~~NPQR}
   -8\,\delta_M^{[N}\hatG^{PQR]})\,\hF_{NPQR}\,,
\end{eqnarray}
$\hat{e}_{M}^{\,~A}$ is the eleven-dimensional bosonic vielbein
(elfbein) and hence
$G_{MN}=\hat{e}_{M}^{\,~A}\hat{e}_N^{\,~B}\eta_{AB}$,
$\hA_{MNP}$ is the bosonic three-form and its field strength
$\hF_{MNPQ}=4\p_{[M} \hA_{NPQ]}$, $\Gamma_A$ is the gamma matrix in
eleven dimensions,\footnote{The gamma matrices satisfy
$\acom{\Gamma_A,\Gamma_B}=2\eta_{AB}$ and the Dirac conjugate for a
general spinor $\psi$ is $\bar{\psi}=i\psi^\dagger \Gamma^0$.}
$\bar{\theta}={}^t\theta\, C={}^t\theta\,\Gamma^0$,
$\Gamma_{A_1A_2\cdots A_n} \equiv
\Gamma_{[A_1}{\Gamma}_{A_2}\cdots\Gamma_{A_n]}\,$,
$\hatG_M\equiv \hat{e}_M^{\,~A}{\Gamma}_A$, and $\hat{\omega}_M^{~AB}$
is the torsion free spin connection
\begin{equation}
 \hat{\omega}_M^{\,~AB}
  =\hat{e}^{[A|N|}(\p_{M}\hat{e}_{N}^{\,~B]}-\p_{N}\hat{e}_{M}^{\,~B]})
  -\frac{1}{2}\,\hat{e}^{[A|N|}\hat{e}^{B]P}(\p_{N}\hat{e}_{P}^{\,~C}
  -\p_{P}\hat{e}_{N}^{\,~C})\hat{e}_{MC}\,.\label{eq:11dsc}
\end{equation}
It is helpful to define the objects with the tangent space indices as
follows
\begin{equation}
 \hat{\Omega}_{A}=\hat{e}_{A}^{~M}\,\hat{\Omega}_{M}\,,\qquad
 \hat{\omega}_A^{\,~BC}=\hat{e}_{A}^{~M}\,\hat{\omega}_M^{\,~BC}\,,
 \qquad\hF_{ABCD}=\hat{e}_{A}^{~M}\hat{e}_{B}^{~N}\hat{e}_{C}^{~P}
	\hat{e}_{D}^{~Q}\,\hF_{MNPQ}\,,\label{eq:11dobj}
\end{equation}
where $\hat{e}_{A}^{~M}$ is the inverse of $\hat{e}_M^{\,~A}$.
In fact, it is easy to work with the tangent space indices rather than
with the target space indices when we calculate the component field
expansion to quadratic order in $\theta$.
Consequently, the supermembrane action in $\theta^2$-order is given
by\footnote{Note that the fermionic coordinate $y$ in
Ref.\cite{MS} corresponds to $-i\sqrt{2}\,\theta$.}
\begin{eqnarray}
 S&=&T\int d^3\sigma\,\left[-\sqrt{-\det \mathbb{G}_{\hi\hj}}
    -\frac{1}{6}\,\epn^{\hi\hj\hk}\hbC_{\hk\hj\hi}\right]\nn
 &=&T\int d^3\sigma\,\left[-\sqrt{-\det G_{\hi\hj}}
    -\frac{1}{6}\,\epn^{\hi\hj\hk}\hat{A}_{\hk\hj\hi}
    -i\sqrt{-\det G_{\hi\hj}}\,\bar{\theta}(1-\hatG_{M2})\hatG^{\hi}
	\hat{D}_{\hi}\theta\right],\qquad\label{eq:theta2S}
\end{eqnarray}
where
\begin{eqnarray}
 \mathbb{G}_{\hi\hj}&=&G_{\hi\hj}+2i\bar{\theta}
	\hatG_{(\hi}\hat{D}_{\hj)}\theta\,,
    \quad\hbC_{\hi\hj\hk}=\hat{A}_{\hi\hj\hk}
    +3i\bar{\theta}\hatG_{[\hi\hj}\hat{D}_{\hk]}\theta\,,\\
 G_{\hi\hj}&=&\p_{\hi}X^M \p_{\hj}X^N \,G_{MN}\,,\quad
    \hat{A}_{\hi\hj\hk}=\p_{\hi}X^M \p_{\hj}X^N
    	\p_{\hk}X^P \hat{A}_{MNP}\,,\\
 \hatG_{\hi}&=&\p_{\hi} X^M \hatG_M\,,\quad
 	\hatG^{\hi}=G^{\hi\hj}\hatG_{\hj}\,,\\
 \hat{\Omega}_{\hi}&=& \p_{\hi} X^M \hat{\Omega}_M\,,\quad
	\hat{D}_{\hi}=\p_{\hi}+\hat{\Omega}_{\hi}\,,\\
 \hatG_{M2}&=&\frac{1}{6}\,\frac{1}{\sqrt{-\det G_{\hi\hj}}}\,
    \epn^{\hi\hj\hk}\hatG_{\hi\hj\hk}\,.
\end{eqnarray}

\section{Double dimensional reduction}\label{S:DDR}
We consider a wrapped supermembrane action compactified on a
2-torus. We shall take the shrinking limit of the 2-torus, or make the
double dimensional reduction\cite{DHIS}, and perform the T-dual
transformation to deduce the $(p,q)$-string action directly from the
supermembrane action (\ref{eq:SMac}) or (\ref{eq:Mac1}).
We take the 10th and 9th directions to compactify on $T^2$,
whose radii are $L_1$ and $L_2$, respectively.
In taking the shrinking volume limit of the 2-torus, we keep the ratio
of the radii finite, or fix the moduli of $T^2$,
\begin{equation}
 \gsb\equiv \frac{L_1}{L_2}~\mbox{: finite.}\quad (L_1,L_2\to0)
 \label{eq:gb}
\end{equation}
Considering the line element on the 2-torus
\begin{equation}
 ds^2_{T^2}=G_{uv}\,dX^udX^v=\Big(G_{99}
    -\frac{(G_{910})^2}{G_{1010}}\Big)(dX^9)^2+G_{1010}
    \Big(dX^{10}+\frac{G_{910}}{G_{1010}}dX^9\Big)^2\,,\label{eq:dst2}
\end{equation}
where $u,v=9,10$, we shall impose that the target space coordinates
satisfy the following boundary conditions \cite{OUY1}
\begin{eqnarray}
 \sqrt{\GO_{1010}}\,X^{10}(\sigma^1,\sigma^2+2\pi)&=&2\pi w_1 L_1 p
    +\sqrt{\GO_{1010}}\,X^{10}(\sigma^1,\sigma^2)\,,\nn
	\sqrt{\GO_{99}-\frac{(\GO_{910})^2}{\GO_{1010}}}\,
	X^{9}(\sigma^1,\sigma^2+2\pi)
  &=&2\pi w_1 L_2 q +\sqrt{\GO_{99}-\frac{(\GO_{910})^2}{\GO_{1010}}}
	\,X^{9}(\sigma^1,\sigma^2)\,,\nn
  \sqrt{\GO_{1010}}\,X^{10}(\sigma^1+2\pi,\sigma^2)&=& 2\pi w_2 L_1 r
    +\sqrt{\GO_{1010}}\,X^{10}(\sigma^1,\sigma^2)\,,\nn
	\sqrt{\GO_{99}-\frac{(\GO_{910})^2}{\GO_{1010}}}\,
	X^{9}(\sigma^1+2\pi,\sigma^2)
  &=& 2\pi w_2 L_2 s +\sqrt{\GO_{99}-\frac{(\GO_{910})^2}{\GO_{1010}}}
	\,X^{9}(\sigma^1,\sigma^2)\,,\label{eq:pq2s}
\end{eqnarray}
where\footnote{We may assume $\Nis>0$ and $w_1>0$ without loss of
generality since we can flip the signs of $(p,q)\to(-p,-q)$ (for
$w_1$) and $(r,s)\to(-r,-s)$ (for $\Nis$) if necessary.
Furthermore, we can see that eq.(\ref{eq:pq}) leads to $(r,s)=n(-q,p)$
($n\in\N$).\label{f:1}}
\begin{equation}
 pr+qs=0\,,\quad ps-qr\equiv\Nis>0\,,\quad (p,q,r,s \in\Z,
	\ w_1\in\N\backslash\{0\}\,,
	\ w_2\in\Z\backslash\{0\})\label{eq:pq}
\end{equation}
and $\GO_{1010}$, $\GO_{99}$ and $\GO_{910}$ stand for the asymptotic
constant values of the metric.\footnote{Precisely speaking,
$\GO_{uv}~(u,v=9,10)$ should satisfy $\p_i\GO_{uv}=0$ and
(\ref{eq:gbLL}) as well.}
Eq.(\ref{eq:pq2s}) can be written by
\begin{eqnarray}
 X^{10}(\sigma^1,\sigma^2) &=& R_1\,(w_1p\sigma^2 + w_2r\sigma^1)
    + Y^1(\sigma^1,\sigma^2)\,,\nn
 X^9(\sigma^1,\sigma^2)&=&R_2\,(w_1q\sigma^2+w_2s\sigma^1)
	+Y^2(\sigma^1,\sigma^2)\,,\label{eq:md2'}
\end{eqnarray}
with
\begin{equation}
 Y^\xi(\sigma^1+2\pi,\sigma^2)=Y^\xi(\sigma^1,\sigma^2+2\pi)
	=Y^\xi(\sigma^1,\sigma^2)\,,\quad(\xi=1,2)
\end{equation}
and\footnote{We shall see $R_1=L_1\,e^{-2\pc/3}$ from
eq.(\ref{eq:KKmetric}) where $\pc$ is the asymptotic value of the type
IIA dilaton background and hence M/IIA-relation, or
11d/IIA-SUGRA-relation, leads to $R_1=\ell_{11}$ (the
eleven-dimensional Planck length).\label{fn:1}}
\begin{equation}
 R_1\equiv \frac{L_1}{\sqrt{\GO_{1010}}}~,\quad R_2\equiv
    \frac{L_2}{\sqrt{\GO_{99}-\frac{(\GO_{910})^2}{\GO_{1010}}}}~.
\end{equation}
The other fields satisfy the periodic boundary conditions,
$X^0(\sigma^1+2\pi,\sigma^2)=X^0(\sigma^1,\sigma^2+2\pi)
=X^0(\sigma^1,\sigma^2)$, etc..
The above expressions represent that the supermembrane is wrapping
$w_1p$-times around one of the two compact directions
(the $X^{10}$-direction) and $w_1q$-times around the other direction
(the $X^9$-direction), or $w_1$-times around $(p,q)$-cycle along the
$\sigma^2$-direction on the worldsheet.
And it is also wrapping $w_2$-times around $(r,s)$-cycle along the
$\sigma^1$-direction.
These two cycles are orthogonal to each other and intersect at least
once. Thus, this wrapped supermembrane is expected to give the
$(p,q)$-string \cite{Sch,OUY1}.
In fact, we shall see below that the $(p,q)$-string comes
out through the double dimensional reduction.

Now that we shall adopt the double dimensional reduction technique
\cite{DHIS} to deduce $(p,q)$-strings. First we determine the
spacetime direction to be aligned with one of the worldvolume
coordinate, or we fix the gauge.
We define $X^y$ and $X^z$ by an SO(2) rotation of the target space,
\begin{equation}
  \left(\begin{array}{@{\,}c@{\,}} X^z \\ X^y \end{array}\right)=\Opq
    \left(\begin{array}{@{\,}c@{\,}}X^{10} \\ X^9 \end{array}\right),
	\label{eq:so2rot}
\end{equation}
where
\begin{equation}
  \Opq =\frac{1}{\cpq}\left(\begin{array}{@{\,}cc@{\,}}
        p & q\\  -q & p \end{array}\right)
	\equiv \left(\begin{array}{@{\,}cc@{\,}}
        \hp & \hq\\  -\hq & \hp \end{array}\right)\in SO(2)\,,
 \quad\cpq\equiv \sqrt{p^2 +q^2}\,. \label{eq:so2mx}
\end{equation}
By using the relations between the eleven-dimensional supergravity
fields and the $S^1$-compactified type IIB ones \cite{BHO,MO}, we have
\begin{equation}
  \sqrt{\frac{\hG_{99}-\frac{(\hG_{910})^2}{\hG_{1010}}}{\hG_{1010}}}
  = e^{-\varphi}\to\sqrt{\frac{\GO_{99}
	-\frac{(\GO_{910})^2}{\GO_{1010}}}{\GO_{1010}}}
	=e^{-\vpc}=\gsb^{-1}=\frac{L_2}{L_1}\,,\label{eq:gbLL}
\end{equation}
where $\varphi$ is the type IIB dilaton background and $\vpc$ is
its asymptotic constant value.
Thus, eq.(\ref{eq:gbLL}) leads to
\begin{equation}
  R_1 = R_2 \equiv R_B\,.\label{eq:RB}
\end{equation}
Then we have
\begin{eqnarray}
  X^z&=&w_1\cpq\,R_B\,\sigma^2
	+\hp\,Y^1(\sigma^{\hi})+\hq\,Y^2(\sigma^{\hi})\,,\nn
  X^y&=&\frac{w_2\,\Nis R_B}{\cpq}\,\sigma^1
	-\hq\,Y^1(\sigma^{\hi})+\hp\,Y^2(\sigma^{\hi})\,.
\end{eqnarray}
The target space metric and the background 3-form field are
transformed under the SO(2) rotation in eq.(\ref{eq:so2rot}) as
($\RM,\RN,\RP,\RQ=0,1,2,\cdots,8,y,z$)
\begin{equation}
 \tG_{\RM\RN} = G_{MN}\,\frac{\p X^M}{\p X^{\RM}}\,
	\frac{\p X^N}{\p X^{\RN}}\,,\quad
   \thA_{\RM\RN\RP}= \hA_{MNP}\,\frac{\p X^M}{\p X^{\RM}}\,
    \frac{\p X^N}{\p X^\RN}\,\frac{\p X^P}{\p X^\RP}\,.
 \label{eq:11dGA'}
\end{equation}
In addition, we shall define the objects in the rotated coordinate
system as follows
\begin{eqnarray}
 &&\thG{}^{\RM}=\thse_{A}^{~\RM}\,{\Gamma}^A\,,\qquad
    \thO_{\RM}=\thse_{\RM}^{~A}\,\hat{\Omega}_{A}\,,\nn
 &&\tho_{\RM}{}^{BC}=\thse_{\RM}{}^A\,\hat{\omega}_A{}^{BC}\,,\quad
    \thF_{\RM\RN\RP\RQ}=\thse_{\RM}{}^A\,\thse_{\RN}{}^B\,
    \thse_{\RP}{}^{C}\,\thse_{\RQ}{}^D\,\hF_{ABCD}\,.\label{eq:11dobj'}
\end{eqnarray}

Let us choose the Kaluza-Klein condition for the target space metric
and the bosonic vielbein in the rotated coordinate system.
A suitable choice is ($\rmu,\rnu=0,1,\cdots,8,y$ and
$\mu,\nu=0,1,\cdots,8$)
\begin{eqnarray}
 \tG_{\RM\RN}&\equiv&e^{-\frac{2}{3}\tphi}
    \left(\begin{array}{@{\,}cc@{\,}}
    \tg_{\rmu\rnu}+e^{2\tphi}\tA_{\rmu}\tA_{\rnu}&
	e^{2\tphi}\tA_{\rmu} \\[10pt]
	e^{2\tphi}\tA_{\rnu} & e^{2\tphi}\end{array}\right)\nn
  &=&\left(\begin{array}{@{\,}ccc@{\,}}
	\frac{1}{\sqrt{\tG_{zz}}}\,\tg_{\mu\nu}
	+\frac{1}{\tG_{zz}}\tG_{\mu z}\tG_{\nu z}&
	\frac{1}{\sqrt{\tG_{zz}}}\,\tg_{\mu y}
	+\frac{1}{\tG_{zz}}\tG_{\mu z}\tG_{yz}& \tG_{\mu z} \\[10pt]
	\frac{1}{\sqrt{\tG_{zz}}}\,\tg_{y\nu}
	+\frac{1}{\tG_{zz}}\tG_{yz}\tG_{\nu z}&
	\frac{1}{\sqrt{\tG_{zz}}}\,\tg_{yy}
	+\frac{1}{\tG_{zz}}\tG_{yz}\tG_{yz}& \tG_{yz} \\[10pt]
	\tG_{\nu z}&\tG_{yz} & \tG_{zz}
	\end{array}\right),\label{eq:KKtG}
\end{eqnarray}
and
\begin{equation}
 \thse_{\RM}{}^A
  =e^{-\frac{\tphi}{3}}\left(\begin{array}{@{\,}cc@{\,}}
    \tse_{\rmu}^{~\hr}& e^{\tphi}\tA_{\rmu}\\[10pt]
    0 & e^{\tphi} \end{array}\right),\quad
 \thse^{~\RM}_{A}
  =e^{\frac{\tphi}{3}}\left(\begin{array}{@{\,}cc@{\,}}
    \tse_{\hr}^{~\rmu} & -\tse_{\hr}^{~\rmu}\tA_{\rmu}\\[10pt]
    0 & e^{-\tphi}\end{array}\right),\label{eq:11dKKte}
\end{equation}
where $\tse_{\rmu}^{~\hr}$ and $\tg_{\rmu\rnu}$ are the vielbein
(zehnbein) and the target space metric in ten dimensions,
respectively,
$\tA_{\rmu}$ is a Kaluza-Klein vector field and $\tphi$ is a
scalar field which is reduced to the type IIA dilaton in the case of
$(p,q)=(1,0)$ in eq.(\ref{eq:so2mx}).

Eq.(\ref{eq:11dKKte}) implies that the spinor $\theta$ should be
rescaled by $e^{-\tphi/6}$ in ten dimensions, which is understood as
follows.
The gamma matrices in eleven dimensions are split into the
ten-dimensional gamma matrices and the rest,
$\{\Gamma_A\}=\{\Gamma_{\hr},\Gamma_{10}\}$.
Let us consider the supersymmetry transformation, $\delta X^{M}=
i\bar{\epn}\,\hatG^M\theta =
i\hat{e}^{~M}_{A}\,\bar{\epn}\,\Gamma^A\theta,\ \delta\theta =\epn$.
Then, we have $\delta X^{\rmu} =i\hat{e}^{~\rmu}_{A}\,\bar{\epn}\,
\Gamma^A\theta=ie^{\tphi/3}\tse^{~\rmu}_{\hr}\,\bar{\epn}\,
\Gamma^{\hr}\theta,\ \delta\theta=\epn$.
Once we impose that the form of the transformation in eleven
dimensions is preserved in ten dimensions, we should rescale the
Majorana spinors $\theta\to e^{-\tphi/6}\,\theta\,$ and $\epn\to
e^{-\tphi/6}\,\epn\,$.
In addition, we also define the following  objects in ten dimensions
\begin{equation}
 \tOmg_{\rmu}\equiv\tse_{\rmu}^{~\hr}\,{\Omega}_{\hr}\,,\quad
 \Omega_{\hr}\equiv e^{-\frac{\tphi}{3}}\, \hat{\Omega}_{\hr}\,,\quad
 \Omega_{10}\equiv e^{-\frac{\tphi}{3}}\,\hat{\Omega}_{10}\,,
 \label{eq:10Omg}
\end{equation}
and these are related to $\{\thO_{\RM}\}=\{\thO_{\rmu},\thO_{z}\}$ in
eq.(\ref{eq:11dobj'}) as
\begin{equation}
 \thO_{\rmu}=\thse_{\rmu}{}^{A}\,\hat{\Omega}_{A}
  =\tOmg_{\rmu}+e^{\tphi}\tA_{\rmu}\,{\Omega}_{10}\,,\quad
 \thO_{z}=\thse_{z}{}^{A}\,\hat{\Omega}_{A}
	=e^{\tphi}\,{\Omega}_{10}\,.
\end{equation}
The supervielbein analog of the Kaluza-Klein condition
(\ref{eq:11dKKte}) is given by ($\tm=(\rmu,\alpha)$)
\begin{equation}
 \thE_{\hat{\RM}}{}^{\hat{A}}=\left(\begin{array}{@{\,}ccc@{\,}}
    \thE_{\tm}{}^{\hr} & \thE_{\tm}{}^a & \thE_{\tm}{}^{10} \\
    \thE_{z}{}^{\hr}&\thE_{z}{}^a &\thE_{z}{}^{10}\end{array}\right)
  =\tilde{\Phi}^{-\frac{1}{3}}\left(\begin{array}{@{\,}ccc@{\,}}
    \tE_{\tm}^{~\hr} & \tE_{\tm}^{~a}+\tilde{C}_{\tm}\tilde{\psi}^a
	& \tilde{\Phi} \tilde{C}_{\tm}\\
    0 & \tilde{\psi}^a & \tilde{\Phi}\end{array}\right),
    \label{eq:11dKKtE}
\end{equation}
which implies
\begin{equation}
 \thse{}_{z}^{~\hr}=0\,,\quad
  \bar{\theta}\Gamma^{\hr}\thO_z\theta=0\,.\quad
  (\,\bar{\theta}\Gamma^{\hr}\Omega_{10}\theta=0\,)
\end{equation}

Now we shall make a (partial) gauge choice of (cf. Ref.\cite{DHIS})
\begin{equation}
  X^z=\frac{L_1w_1\cpq}{\sqrt{\GO_{1010}}}\,\sigma^2\,
	\equiv {\zl}\,\sigma^2,\label{eq:DDR0}
\end{equation}
or the $X^z$-direction is aligned with one of the space direction
$\sigma^2$ of the worldvolume. Then the dimensional reduction is
achieved by imposing the following conditions on the target superspace
coordinates and the background fields,
\begin{eqnarray}
  &&\frac{\p}{\p \sigma^2}\,Z^{\tm}=0\,,\label{eq:DDR1}\\
 && \frac{\p}{\p X^z}\,\tG_{\RM\RN}
    = \frac{\p}{\p X^z}\,\thA_{\RM\RN\RP} =0\,.\label{eq:DDR2}
\end{eqnarray}
Thus the induced metric on the worldvolume is given by \cite{DHIS}
($i,j=0,1$)
\begin{equation}
  \hat\gamma_{\hi\hj}=\hPi_{\hi}^{~A} \hPi_{\hj}^{~B}\,\eta_{AB}
  =\Phi'^{-\frac{2}{3}}\left(\begin{array}{@{\,}cc@{\,}}
       \gamma_{ij}+ \Phi'^2 C_iC_j & \Phi'^2 C_i \\[10pt]
       \Phi'^2 C_j & \Phi'^2 \end{array}\right)\,,
\end{equation}
where
\begin{eqnarray}
 \Phi'^{\frac{4}{3}}&=&{\zl}^2\,\tilde{\Phi}^{\frac{4}{3}}\,,\nn
 \Phi'^{\frac{4}{3}}C_i&=&{\zl}\,\tilde{\Phi}^{\frac{4}{3}}\,
	\p_i X^{\tm}\tilde{C}_{\tm}\,,\nn
  \gamma_{ij}&=&{\zl}\,\tilde{\Pi}_i^{~\hr}\tilde{\Pi}_j^{~\hs}\,
    \eta_{\hr\hs}\,,\qquad
      (\tilde{\Pi}_i^{~\hr}\equiv\p_i Z^{\tm}\tE_{\tm}^{~\hr})
\end{eqnarray}
and (up to quadratic order in $\theta$ with the fermionic background
fields being zero)\footnote{Note that we have taken the rescaling of
$\theta$, $\theta \to e^{-\tphi/6}\,\theta$.}
\begin{eqnarray}
 \tilde{\Phi}^{\frac{1}{3}}&=&(\thE_z^{~10})^{\frac{1}{2}}
    =e^{\frac{\tphi}{3}}\,\Bigl(1+\frac{i}{2}\,\bar{\theta}
		\Gamma_{10}\Omega_{10}\theta\Bigr)\,,\nn
 \tE_{\rmu}{}^{\hr}&=&\tilde{\Phi}^{\frac{1}{3}}\thE_{\rmu}^{~\hr}
    =\Bigl(1+\frac{i}{2}\,\bar{\theta}\Gamma_{10}
	\Omega_{10}\theta\Bigr)\,\tse_{\rmu}^{~\hr}
	+i\bar{\theta}\Gamma^{\hr}\tOmg_{\rmu} \theta\,,\nn
 \tE_{\alpha}{}^{\hr}&=&\tilde{\Phi}^{\frac{1}{3}}\thE_\alpha^{~\hr}
    =-e^{\frac{\tphi}{6}}\,i(\bar{\theta}\Gamma^{\hr})_{\alpha}\,,\nn
 \tilde{\Pi}_i^{~\hr}&=&\p_i Z^{\tm}\tE_{\tm}{}^{\hr}
 	=\p_i X^{\rmu}\,\Bigl\{\tse_{\rmu}^{~\hr}\Bigl(1
	+\frac{i}{2}\,\bar{\theta}\Gamma_{10}\Omega_{10}\theta\Bigr)
	+i\bar{\theta}\Gamma^{\hr}\tOmg_{\rmu}\theta\Bigr\}
	+i\bar{\theta}\Gamma^{\hr}\p_i \theta\,.
\end{eqnarray}
We have
\begin{equation}
  \sqrt{-\det\hat\gamma_{\hi\hj}}=\sqrt{\mathstrut-\det\gamma_{ij}}~.
\end{equation}
Thus, by the double dimensional reduction of
eqs.(\ref{eq:DDR0})-(\ref{eq:DDR2}), the supermembrane action
(\ref{eq:Mac1}) is reduced to
\begin{eqnarray}
 S_{ddr}=2\pi T\int\!d\sigma^0\!\int_0^{2\pi}
   \hspace{-1ex}d\sigma^1\,{\zl}\Biggl[-\sqrt{-\det\tbG_{ij}}\,
	-\frac{1}{2}\,\epn^{ij}\tbB_{ji}\Biggr]\,,\label{eq:Macr2}
\end{eqnarray}
where
\begin{eqnarray}
 \tbG_{ij}&=&\tg_{ij}+\tQ_{ij}
	+2i\bar{\theta}\Gamma_{(i}\p_{j)}\theta\,,\\
 \tbB_{ij}&=&\tA_{ij}-\tP_{ij}-2i\bar{\theta}
	\Gamma_{[i}\Gamma^{10}\p_{j]}\theta\,,
\end{eqnarray}
the indices $i,j$ of $\tg_{ij}$, $\tQ_{ij}$, etc. mean
\begin{equation}
  \tg_{ij}\equiv\p_iX^{\rmu}\p_jX^{\rnu}\tg_{\rmu\rnu}\,,\quad
  \tQ_{ij}\equiv\p_iX^{\rmu}\p_jX^{\rnu}\tQ_{\rmu\rnu}\,,\quad
  \Gamma_i\equiv\p_iX^{\rmu}\tse_{\rmu}^{~\hr}\Gamma_{\hr}\,,
	\quad\mbox{etc.}\,,
\end{equation}
and
\begin{eqnarray}
 &&\tA_{\rmu\rnu}=\thA_{\rmu\rnu z}\,,\quad
    \tQ_{\rmu\rnu}=\tse_{\rmu}^{~\hr}\tse_{\rnu}^{~\hs}
      Q_{\hr\hs}\,,\quad\tP_{\rmu\rnu}=\tse_{\rmu}^{~\hr}
	\tse_{\rnu}^{~\hs}P_{\hr\hs}\,,\nn
 &&Q_{\hr\hs}=i\eta_{\hr\hs}\bar{\theta}\Gamma_{10}\Omega_{10}\,
   \theta +2i\bar{\theta} \Gamma_{(\hr} \Omega_{\hs)}\theta\,,
  \quad P_{\hr\hs}=-i\bar{\theta}\Gamma_{\hr\hs}{\Omega}_{10}\theta
   -2i\bar{\theta}\Gamma_{10}\Gamma_{[\hr}\Omega_{\hs]}\theta\,.
   \label{eq:IIAQP}
\end{eqnarray}
This reduced action (\ref{eq:Macr2}) naturally inherits
$\kappa$-symmetry \cite{DHIS}.
In fact, eq.(\ref{eq:K11}) leads to the transformation law which
leaves the action (\ref{eq:Macr2}) of order up to quadratic in
$\theta$ invariant
\begin{equation}
\dkp\theta=(1+\Gamma_{F})\kappa\,,\quad
 \dkp X^{\rmu}=-i\bar{\theta}\Gamma^{\rmu}(1+\Gamma_{F})
    \kappa\,, \quad \dkp\Phi_{bg}
	=\dkp X^{\rmu}\p_{\rmu}\Phi_{bg}\,,\label{eq:Akappa}
\end{equation}
where $\Phi_{bg}$ stands for a general field of supergravity
background and
\begin{equation}
 \Gamma_F=\frac{1}{2}\,\frac{1}{\sqrt{-\det\tg_{ij}}}\,
	\epn^{ij}\Gamma_{ij}\Gamma^{10}\,.\label{eq:GF}
\end{equation}
Similarly, the SUSY transformation, which leaves the action
(\ref{eq:Macr2}) invariant, is given by
\begin{eqnarray}
 \delta_\epn \theta=\epn\,,\quad
 \delta_\epn X^{\rmu}=i\bar{\epn}\Gamma^{\rmu}\theta \,,\quad
 \delta_\epn\Phi_{bg}=\delta_{\epn}X^{\rmu}\p_{\rmu}\Phi_{bg}\,.
 \label{eq:2aSS}
\end{eqnarray}

By introducing the worldsheet metric $\tgamma_{ij}$,
eq.(\ref{eq:Macr2}) can be rewritten in the Polyakov form as usual
\begin{eqnarray}
 S_{ddr}&=&\frac{2\pi T}{2}\int\!d\sigma^0\!\int_0^{2\pi}
   \hspace{-1ex}d\sigma^1\,{\zl}\Biggl[-\sqrt{-\tgamma}\,
    \tgamma^{ij}\,(\p_iX^{\rmu}\p_jX^{\rnu}\tE_{\rmu\rnu}
	+2\p_iX^{\rmu}\tilde{\calG}_{j\rmu})\nn
  &&{}\hspace{23ex}+\epn^{ij}(\p_i X^{\rmu}\p_j X^{\rnu}
   \tilde{B}_{\rmu\rnu}-2\p_i X^{\rmu}\tilde{\calB}_{j\rmu})
   \Biggr]\,,\label{eq:Macr}
\end{eqnarray}
where
\begin{equation}
  \tE_{\rmu\rnu}=\tg_{\rmu\rnu}+\tQ_{\rmu\rnu}\,,\quad
  \tilde{B}_{\rmu\rnu}=\tA_{\rmu\rnu}-\tP_{\rmu\rnu}\,,\label{eq:SEB}
\end{equation}
are the super metric and the super 2-form, respectively, and
\begin{eqnarray}
 &&\tilde{\calG}_{j\rmu}=\tse_{\rmu}^{~\hr}\calG_{j\hr}\,,\quad
	\tilde{\calB}_{j\rmu}=\tse_{\rmu}^{~\hr} \calB_{j\hr}\,,\nn
 && \calG_{j \hr}=i\bar{\theta} \Gamma_{\hr}\p_j\theta\,,
    \quad\calB_{j \hr}=i\bar{\theta}\Gamma_{\hr}
    \Gamma_{10}\,\p_j\theta\,.\label{eq:IIAQP'}
\end{eqnarray}
As was pointed out in \cite{DHIS}, this action (\ref{eq:Macr}) has
conformal invariance.
When we consider the T-dual transformation between type IIA/IIB
superstring theories, it is, in fact, convenient to work with the
action of the Polyakov type. Eq.(\ref{eq:Macr}) can be written by
\begin{equation}
 S_{ddr}=\frac{2\pi T}{2}\int\!d\sigma^0\!\int_0^{2\pi}
   \hspace{-1ex}d\sigma^1\,{\zl}\Biggl[-\sqrt{-\tgamma}\,
    \tgamma^{ij}\p_iX^{\rmu}\p_jX^{\rnu}\tbG_{\rmu\rnu}
    +\epn^{ij}\p_i X^{\rmu}\p_j X^{\rnu}\tbB_{\rmu\rnu}\Biggr]\,,
	\label{eq:Macr0}
\end{equation}
where\footnote{It is understood that $\p_{\rmu}\theta$ is always
associated with $\p_iX^{\rmu}$ in the action to make it a derivative
term w.r.t.\ a worldsheet coordinate, $\p_i\theta$.}
\begin{equation}
  \tbG_{\rmu\rnu}=\tg_{\rmu\rnu}+\tQ_{\rmu\rnu}
	+2i\bar{\theta}\Gamma_{(\rmu}\p_{\rnu)}\theta\,,\quad
  \tbB_{\rmu\rnu}=\tA_{\rmu\rnu}-\tP_{\rmu\rnu}
	-2i\bar{\theta}\Gamma_{[\rmu}\Gamma_{10}\p_{\rnu]}\theta\,.
\end{equation}
 
We shall give $Q_{\hr\hs}$ and $P_{\hr\hs}$ in eq.(\ref{eq:IIAQP})
more explicitly with the background fields.
As we noted, it is easy to work out with the tangent space indices
rather than with the target space indices.
Hence we shall calculate the decomposition of the 4-form field
strength $\hF_{ABCD}$ and the spin connection $\hat{\omega}_A^{~BC}$
in eq.(\ref{eq:11dobj'}) under the Kaluza-Klein condition
(\ref{eq:11dKKte}).
The decomposition of the 4-form field strength
\begin{equation}
  \{\hF_{ABCD}\}=\{\hF_{\hrstu},\hF_{\hrst 10}\}\,,
\end{equation}
is
\begin{equation}
 \hF_{\hrstu}=e^{\frac{4}{3}\tphi}\tse_{\hr}^{~\rmu}
    \tse_{\hs}^{~\rnu}\tse_{\hatt}^{~\rrho}\tse_{\hu}^{~\rsigma}
 \tF_{\rmu\rnu\rrho\rsigma}\equiv
   e^{\frac{4}{3}\tphi}\tF_{\hrstu}\,,
 \quad\hF_{\hrst 10}=e^{\frac{1}{3}\tphi}\,\tse_{\hr}^{~\rmu}
    \tse_{\hs}^{~\rnu}\tse_{\hatt}^{~\rrho}\tH_{\rmu\rnu\rrho}\equiv
   e^{\frac{1}{3}\tphi} \tH_{\hrst}\,,\label{eq:10dFH}
\end{equation}
where
\begin{equation}
 \tF_{\rmu\rnu\rrho\rsigma}\equiv\thF_{\rmu\rnu\rrho\rsigma}
    +4\tA_{[\rmu}\tH_{\rnu \rrho \rsigma]}\,,\quad
 \tH_{\rmu\rnu\rrho}\equiv\thF_{\rmu\rnu\rrho z}\,.\label{eq:10dFH'}
\end{equation}
Similarly, the spin connection
\begin{equation}
 \{\hat{\omega}_{A}^{~BC}\}=\{\hat{\omega}_{\hr}^{~\hs\hatt}\,,
	\hat{\omega}_{\hr}^{~\hs 10}\,,\hat{\omega}_{10}^{~\,\hr\hs}\,,
	\hat{\omega}_{10}^{~\,\hr 10}\}\,,
\end{equation}
is given by
\begin{eqnarray}
 &&\hat{\omega}_{\hr}^{~\hs \hatt}=
    e^{\frac{1}{3}\tphi}\,\Bigl(\tomg_{\hr}^{~\hs\hatt}
      -\frac{2}{3}\,\delta_{\hr}^{~[\hs}\tse^{\hatt]\rnu}
      \p_{\rnu}\tphi\Bigr)\,,\qquad
  \hat{\omega}_{\hr}^{~\hs 10}=\frac{1}{2}\,
    e^{\frac{4}{3}\tphi}\,\tse_{\hr}^{~\rmu}\tse^{\hs\rnu}
    \tF_{\rmu \rnu}\,,\nn
 &&\hat{\omega}_{10}^{~\,\hr\hs}=-\frac{1}{2}\,e^{\frac{4}{3}\tphi}\,
      \tse^{\hr\rrho}\tse^{\hs\rsigma}\tF_{\rrho\rsigma}\,,\qquad
  \hat{\omega}_{10}^{~\,\hr 10}=-\frac{2}{3}\,e^{\frac{1}{3}\tphi}\,
      \tse^{\hr\rnu}\p_{\rnu}\tphi\,, \label{eq:10dsc}
\end{eqnarray}
where $\tomg_{\hr}^{~\hs\hatt}
=\tse_{\hr}^{~\rmu}\tomg_{\rmu}^{~\hs\hatt}$,
$\tomg_{\rmu}^{~\hr\hs}$ is the torsion free spin connection,
which is made of $\tse_{\rmu}^{~\hr}$, and
\begin{equation}
  \tF_{\rmu\rnu}=2\p_{[\rmu}\tA_{\rnu]} \label{eq:10dF2}
\end{equation}
is the field strength of the Kaluza-Klein vector field in
eq.(\ref{eq:11dKKte}).

Then, $\{\hat{\Omega}_A\}=\{\hat{\Omega}_{\hr},\hat{\Omega}_{10}\}$ in
eq.(\ref{eq:11dobj}) is calculated explicitly by using
(\ref{eq:10dFH}), (\ref{eq:10dsc}) and (\ref{eq:10dF2}), and
hence we have (cf. Appendix \ref{S:Bisp})
\begin{eqnarray}
 Q_{\hr\hs}&=&Q_{\hr\hs}^{(2)}-Q_{\hr\hs}^{(4)}
    +\frac{i}{4}\,s^{IJ}\bar{\theta}_I
    \Gamma_{(\hr}^{~~\hatt\hu}\theta_J \tH_{\hs)\hatt\hu}
    +\frac{i}{2}\,\bar{\theta}_I\Gamma_{(\hr}\Gamma_{|\hatt\hu|}
	\theta_I\tomg_{\hs)}^{~\hatt \hu}\,,\nn
 P_{\hr\hs}&=&P_{\hr\hs}^{(2)}-P_{\hr\hs}^{(4)}
    +\frac{i}{4}\,\bar{\theta}_I
    \Gamma_{[\hr}^{~~\hu\hv}\theta_I\tH_{\hs]\hu\hv}
    +\frac{i}{2}\,s^{IJ}\bar{\theta}_I\Gamma_{[\hr}\Gamma_{|\hatt\hu|}
    \theta_J\tomg_{\hs]}^{~\hatt \hu}\,,\nn
 \calG_{i\hr}&=&i\bar{\theta}_I\Gamma_{\hr}\p_i\theta_I\,,\nn
 \calB_{i\hr}&=&is^{IJ}\bar{\theta}_I\Gamma_{\hr}\p_i \theta_J\,,
 \label{eq:IIAQP''}
\end{eqnarray}
where $I,J=+,-$, $s^{++}-1=s^{--}+1=s^{+-}=s^{-+}=0$,
\begin{eqnarray}
 Q^{(n)}_{\hr\hs}&=&\frac{i}{2n!}\,e^{\tphi}\,\bar{\theta}_+
	\Gamma_{(\hr}\Gamma^{\hr_1\cdots\hr_n}\Gamma_{\hs)}\theta_-\,
	\tF_{\hr_1\cdots\hr_n}\,,\quad(n=2,4)\nn
 P^{(n)}_{\hr\hs}&=&\frac{i}{2n!}\,e^{\tphi}\,\bar{\theta}_+
	\Gamma_{[\hr}\Gamma^{\hr_1\cdots\hr_n}\Gamma_{\hs]}\theta_-\,
	\tF_{\hr_1\cdots\hr_n}\,,\quad(n=2,4)
\end{eqnarray}
and $\theta_{\pm}$ are the 16-component positive and negative chiral
spinors, respectively
\begin{equation}
  \theta_\pm=\Gamma_{\pm}\,\theta\,,\quad
	\Gamma_{\pm}=\frac{1}{2}\,(1\pm\Gamma_{10})\,.\label{eq:chsp}
\end{equation}
Then, eq.(\ref{eq:Macr2}) can also be written by (see section
\ref{S:Cal2a})
\begin{eqnarray}
 S_{ddr}=2\pi T\int\!d^2 \sigma\,{\zl}\Biggl[-\sqrt{-\tg}\,
    -\frac{1}{2}\,\epn^{ij}\tA_{ji} -2i\sqrt{-\tg}\,\bar{\theta}
	P_{(-)}\Gamma^i\nabla_i\theta\Biggr]\,,\label{eq:IIA}
\end{eqnarray}
where $\tg=\det\tg_{ij}$ and
\begin{eqnarray}
  P_{(-)}&=&\frac{1}{2}\Bigl(1-\frac{1}{2\sqrt{-\tg}}\,
	\epn^{ij}\Gamma_{ij}\Gamma^{10}\Bigr)\,.\\
  \nabla_i&=&D_i+\frac{1}{8}\,e^{\tphi}\Bigl(
      \frac{1}{2!}\,\Gamma^{\hr\hs}\Gamma^{10}\tF_{\hr\hs}
      -\frac{1}{4!}\,\Gamma^{\hrstu}\tF_{\hrstu}\Bigr)\Gamma_i
	+\frac{1}{8}\,\Gamma^{\hr\hs}\Gamma^{10}\tH_{i\hr\hs}\,,\\
   D_i&=&\p_i+\frac{1}{4}\,\p_iX^{\rmu}\Gamma_{\hr\hs}\,
	\tomg_{\rmu}^{~\hr\hs}\,,\\
 \tH_{i\hr\hs}&=&\p_iX^{\rmu}\tse_{\rmu}^{~\hatt}\tH_{\hatt\hr\hs}\,,
 \quad\Gamma_i=\p_i X^{\rmu} \Gamma_{\rmu}\,,\quad
 \Gamma^i=\tg^{ij}\Gamma_j\,.
\end{eqnarray}

\section{\boldmath$(p,q)$-string from wrapped
supermembrane}\label{S:C}
In this section we derive the $(p,q)$-string action from the reduced
supermembrane action in eq.(\ref{eq:Macr}).
The action has an abelian isometry associated with the other
compactified $X^y$-direction, and we can make a dual transformation as
is the case with sigma models.
Introducing a variable $\XB^9$, which will be seen to be dual to
$X^y$, eq.(\ref{eq:Macr}) (or (\ref{eq:Macr0})) can be rewritten in
a classically equivalent form
\begin{eqnarray}
 S_{ddr}&=&-\frac{2\pi T}{2}\int\!d\sigma^0\!\int_0^{2\pi}
    \hspace{-1ex}d\sigma^1\,{\zl}\,\Biggl[\sqrt{-\tgamma}\,
      \tgamma^{ij}\,(\p_iX^{\mu}\p_jX^{\nu}\,\tbG_{\mu\nu}
      +2\p_iX^{\mu}\Xi_j\,\tbG_{\mu y}+\Xi_i\Xi_j\,\tbG_{yy})\nn
  &&{}\quad-\epn^{ij}(\,\p_iX^{\mu}\p_jX^{\nu}\,\tbB_{\mu\nu}
      +2\p_iX^{\mu} \Xi_j\,\tbB_{\mu y})
      -2\epn^{ij}\XB^9\p_i\Xi_j \Biggr]\,,\label{eq:d-Macr}
\end{eqnarray}
since the variation w.r.t.\ $\XB^9$ leads to $\epn^{ij}\p_i\Xi_j=0$
or $\Xi_j=\p_j X^y$ and hence eq.(\ref{eq:Macr}) can be
reproduced.\footnote{We assume that the background fields are
independent of $\XB^9$ in eq.(\ref{eq:d-Macr}).}
On the other hand, assuming that all the fields are independent of
$\Xi_j$ (or $X^y$), the variation w.r.t.\ $\Xi_i$ leads to
\begin{equation}
  \Xi_i=\frac{\tgamma_{ij}\epn^{kj}}{\tbG_{yy}\sqrt{-\tgamma}}\,
    (\p_k X^\mu\tbB_{\mu y}-\p_k\XB^9)
    -\frac{\tbG_{\mu y}}{\tbG_{yy}}\,\p_i X^\mu\,.\label{eq:d-Y}
\end{equation}
Plugging eq.(\ref{eq:d-Y}) into eq.(\ref{eq:d-Macr}) we have
\begin{eqnarray}
 S_{ddr}&=&-\frac{2\pi T}{2}\int\!d\sigma^0\!\int_0^{2\pi}
   \hspace{-1ex}d\sigma^1\,{\zl}\Biggl[\sqrt{-\tgamma}\,
    \tgamma^{ij}\p_i\XB^{\hmu}\p_j\XB^{\hnu}\ydbG_{\hmu\hnu}
    -\epn^{ij}\p_i\XB^{\hmu}\p_j\XB^{\hnu}\ydbB_{\hmu\hnu}\Biggr]\nn
 &=&-\frac{2\pi T}{2}\int\!d^2\sigma\,{\zl}\,\Biggl[\sqrt{-\tgamma}\,
      \tgamma^{ij}\,(\p_i\XB^{\hmu}\p_j\XB^{\hnu}\ydE_{\hmu\hnu}
      +2\p_i\XB^{\hmu}\ydcalG_{j\hmu})\nn
 &&\hspace{10ex}{}
 	-\epn^{ij}(\p_i\XB^{\hmu}\p_j\XB^{\hnu}\ydB_{\hmu\hnu}
	-2\p_i\XB^{\hmu}\ydcalB_{j\hmu})\Biggr]\,,\label{eq:ddrac}
\end{eqnarray}
where we have defined $\XB^{\hmu}=(\XB^\mu,\XB^9)=(X^\mu,\XB^9)$ and
the ten-dimensional dual fields
\begin{eqnarray}
 &&\ydbG_{\mu\nu}=\tbG_{\mu\nu}+\tbG_{yy}^{-1}
	(\tbB_{\mu y}\tbB_{\nu y}-\tbG_{\mu y}\tbG_{\nu y})\,,
    \quad\ydbG_{\mu 9}=-\tbG_{yy}^{-1}\tbB_{\mu y}\,,
    \quad\ydbG_{99}=\tbG_{yy}^{-1}\,,\nn
 &&\ydbB_{\mu\nu}=\tbB_{\mu\nu}-2\tbG_{yy}^{-1}\tbB_{[\mu|y|}
    \tbG_{\nu]y}\,,\quad \ydbB_{\mu9}=-\tbG_{yy}^{-1}
	\tbG_{\mu y}\,,\label{eq:Tdall0}
\end{eqnarray}
or
\begin{eqnarray}
 &&\ydE_{\mu\nu}=\tE_{\mu\nu}
  +\tE_{yy}^{-1}(\tilde{B}_{\mu y}\tilde{B}_{\nu y}
  -\tE_{\mu y}\tE_{\nu y})\,,\quad
  \ydE_{\mu 9}= -\tE_{yy}^{-1}\tilde{B}_{\mu y}\,,
  \quad \ydE_{99}= \tE_{yy}^{-1}\,,\nn
 &&\ydB_{\mu\nu}=\tilde{B}_{\mu\nu}-\tE_{yy}^{-1}
    (\tilde{B}_{\mu y}\tE_{\nu y}-\tilde{B}_{\nu y}\tE_{\mu y})\,,
    \quad\ydB_{\mu 9}=-\tE_{yy}^{-1}\tE_{\mu y}\,,\nn
 &&\ydcalG_{j\mu}=\tilde{\calG}_{j\mu}+\tE_{yy}^{-1}
   (\tilde{B}_{\mu y}\tilde{\calB}_{jy}
   -\tE_{\mu y}\tilde{\calG}_{jy})\,,
  \qquad\ydcalG_{j9}=-\tE_{yy}^{-1}\tilde{\calB}_{j y}\,,\nn
 &&\ydcalB_{j\mu}=\tilde{\calB}_{j\mu}
  +\tE_{yy}^{-1}(\tilde{B}_{\mu y}\tilde{\calG}_{jy}
  -\tilde{\calB}_{jy}\tE_{\mu y})\,,\qquad
   \ydcalB_{j9}=-\tE_{yy}^{-1}\tilde{\calG}_{jy}\,.\label{eq:Tdall}
\end{eqnarray}

\subsection{$\theta^0$-order action}
The $\theta^0$-order part of eq.(\ref{eq:Tdall0}) is given by
\begin{eqnarray}
 &&\ydg_{\mu\nu}=\tg_{\mu\nu}+\tg_{yy}^{-1}(\tA_{\mu y}\tA_{\nu y}
	-\tg_{\mu y}\tg_{\nu y})\,,
    \quad \ydg_{\mu9}=-\tg_{yy}^{-1}\tA_{\mu y}\,,
    \quad \ydg_{99}=\tg_{yy}^{-1}\,,\nn
 &&\ydA_{\mu\nu}=\tA_{\mu\nu}-2\tg_{yy}^{-1}\tA_{[\mu|y|}
	\tg_{\nu] y}\,,\qquad
   \ydA_{\mu 9}=-\tg_{yy}^{-1}\tg_{\mu y}\,.\label{eq:Td0}
\end{eqnarray}
Note that in the case of $(p,q)=(1,0)$ in eq.(\ref{eq:so2mx}),
eq.(\ref{eq:Td0}) is reduced to the ordinary Buscher's T-dual rule for
the NSNS sector \cite{Bus}.
However, since we have rotated the target space coordinates in
eq.(\ref{eq:so2rot}) by the SO(2) matrix in eq.(\ref{eq:so2mx}),
the T-dual transformation rule (\ref{eq:Td0}) of order zero in
$\theta$ includes not only the NSNS fields but also the RR fields, or
the RR 2-form, which will been seen below.

Now that we consider T-dual for the background fields in
eq.(\ref{eq:Macr}) (or eq.(\ref{eq:ddrac})).
Since we regard $X^{10}$ ({\sl not} $X^{z}$) as the 11th direction,
we should take T-dual along the $X^9$-direction ({\sl not}
$X^y$-direction) to transform type IIA superstring theory to type IIB
superstring theory.
Then we can rewrite the background fields in terms of those of the
type IIB supergravity as follows (cf. Appendix
\ref{S:R}),\footnote{Eq.(\ref{eq:Td0'}) leads to
\begin{eqnarray}
 &&\ydg_{\mu\nu}=\Dpq\jmath_{\mu\nu}\,,
    \quad\ydg_{\mu9}=\Dpq\jmath_{9y}\,,
    \quad\ydg_{99}=\Dpq\jmath_{99}\,,\nn
 &&\ydA_{\mu\nu}=\Dpq\Bpq_{\mu\nu}\,,
   \qquad\ydA_{\mu9}=-\Dpq\Bpq_{9\mu}\,.\label{eq:Td1}
\end{eqnarray}}
\begin{eqnarray}
 \tg_{\mu\nu}&=&\Dpq\,\Bigl(\jmath_{\mu\nu}
	-\frac{\jmath_{9\mu}\jmath_{9\nu}
	-\Bpq_{9\mu}\Bpq_{9\nu}}{\jmath_{99}}\Bigr)\,,\nn
 \tg_{\mu y}&=&\frac{\Bpq_{9\mu}}{\jmath_{99}}\,,\nn
 \tg_{yy}&=&\frac{1}{\Dpq\,\jmath_{99}}\,,\nn
 \thA_{\mu\nu z}\,(=\hq \hA_{\mu\nu9}+\hp A_{\mu\nu10})&=&
    \Dpq\,\Bigl(\Bpq_{\mu\nu}
	+\frac{2\Bpq_{9[\mu}\jmath_{\nu]9}}{\jmath_{99}}\Bigr)\,,\nn
 \thA_{\mu yz}\,(=\hA_{\mu910})
 &=&-\frac{\jmath_{9\mu}}{\jmath_{99}}\,, \label{eq:Td0'}
\end{eqnarray}
where
\begin{equation}
  \Dpq=\sqrt{(\hp+\hq l)^2 +e^{-2\varphi}\hq^2}\,,\label{eq:Dpq}
\end{equation}
 $\BNS_{\mu\nu}$ and $\BR_{\mu\nu}$ are the NSNS and the RR
second-rank antisymmetric tensors, respectively, $\jmath_{\hmu\hnu}$
is the metric in type IIB supergravity, $l=\hG_{910}/\hG_{1010}=A_9$
and
\begin{equation}
  \Bpq_{\hmu\hnu}\equiv \frac{\hp\BNS_{\hmu\hnu}
    +\hq\BR_{\hmu\hnu}}{\Dpq}\,.\label{eq:Bpq}
\end{equation}
Then, plugging these equations into eq.(\ref{eq:Td0}),
the $\theta^0$-order part of the action in
eq.(\ref{eq:ddrac}) is reduced to \cite{OUY1}
\begin{equation}
  S_{ddr}\big|_{\theta^0}=-\frac{2\pi T}{2}\int\!d\sigma^0\!
	\int_0^{2\pi}\hspace{-1ex}d\sigma^1\,{\zl}\,\Dpq
  \Biggl[\sqrt{-\tgamma}\,\tgamma^{ij}\,
	\p_i\XB^{\hmu}\p_j\XB^{\hnu}\jmath_{\hmu\hnu}
	-\epn^{ij}\p_i\XB^{\hmu}\p_j\XB^{\hnu}\Bpq_{\hmu\hnu}
	\Biggr]\,.\label{eq:pqac0}
\end{equation}
We regard $X^{10}$ as the 11th direction, therefore the type IIA
string
tension $T_s$ is given by $2\pi L_1T/\sqrt{\GO_{1010}}$ \cite{Sch}
since the eleven-dimensional metric $\hG_{MN}$ is converted to the
type IIA metric $g_{\hmu\hnu}$ by the relation
$\hG_{\hmu\hnu}=g_{\hmu\hnu}/\sqrt{\hG_{1010}}\,$.
Furthermore, putting the background fields $\{l,\varphi\}$ to be the
asymptotic constant values $\{l_0,\varphi_0\}$, respectively and hence
$e^{\vpc}=\gIIB$, we have
\begin{equation}
  2\pi T {\zl}\Dpq\to
    w_1\,T_s\sqrt{(p+q\lc)^2+e^{-2\vpc}q^2}\equiv w_1\,T_{pq}\,,
\end{equation}
where $T_{pq}$ is the tension of a $(p,q)$-string in type IIB
superstring theory \cite{Sch}.
We can see that both the NSNS and the RR antisymmetric tensors
have coupled to $\XB^{\hmu}$ in eq.(\ref{eq:pqac0}),
which implies that the reduced action (\ref{eq:pqac0}) is, in fact,
that of $(p,q)$-strings.
Note that $w_1$ is just the number of copies of the resulting
$(p,q)$-string. If we allow $q$ to be zero and take
$(p,q,r,s)=(1,0,0,1)$, we have the fundamental strings in type IIB
superstring theory. On the other hand, $(p,q,r,s)=(0,1,1,0)$ leads to
the strings which couple minimally with the RR B-field, i.e., the
D-strings.

\subsection{$\theta^2$-order action}
We next proceed to consider the $\theta^2$-order terms in
eq.(\ref{eq:Tdall}),
\begin{eqnarray}
 &&\ydQ_{\mu\nu}=\tQ_{\mu\nu}-2\tg_{yy}^{-1}
	(\tA_{(\mu|y|}\tP_{\nu) y}+\tg_{(\mu|y|}\tQ_{\nu)y})
	-\tg_{yy}^{-2}\tQ_{yy}(\tA_{\mu y}\tA_{\nu y}
	-\tg_{\mu y}\tg_{\nu y})\,,\nn
 &&\ydQ_{\mu 9}=-\tg_{yy}^{-1}(-\tP_{\mu y}
 	-\tg_{yy}^{-1}\tQ_{yy}\tA_{\mu y})\,,\qquad
	\ydQ_{99}=-\tg_{yy}^{-2}\tQ_{yy}\,, \nn
 &&\ydP_{\mu\nu}=\tP_{\mu\nu}-2\tg_{yy}^{-1}
	(-\tA_{[\mu|y|}\tQ_{\nu] y}
 	+\tP_{[\mu|y|}\tg_{\nu]y}+\tg_{yy}^{-1}
	\tQ_{yy}\tA_{[\mu|y|}\tg_{\nu] y})\,, \nn
 &&\ydP_{\mu9}=\tg_{yy}^{-1}(\tQ_{\mu y}
 	-\tg_{yy}^{-1}\tQ_{yy}\tg_{\mu y})\,,\nn
 &&\ydcalG_{j\mu}=\tcalG_{j\mu}+\tg_{yy}^{-1}
    (\tA_{\mu y}\tcalB_{jy}-\tcalG_{jy}\tg_{\mu y})\,,\qquad
	\ydcalG_{j9}=-\tg_{yy}^{-1}\tcalB_{jy}\nn
 &&\ydcalB_{j\mu}=\tcalB_{j\mu}+\tg_{yy}^{-1}(
      \tA_{\mu y}\tcalG_{jy}-\tcalB_{jy}\tg_{\mu y})\,,\qquad
    \ydcalB_{j9}=-\tg_{yy}^{-1}\tcalG_{jy}\,.\label{eq:Td2}
\end{eqnarray}
We shall plug eq.(\ref{eq:IIAQP''}) into eq.(\ref{eq:Td2}) and
rewrite them by using the type IIB superstring variables.
Note that $\tQ_{\mu\nu}, \tP_{\mu\nu}$, etc. include the vielbein
$\tse_{\rmu}^{~\hr}$, which is transformed under T-dual
according to eq.(\ref{eq:Td0'}), and hence we need the T-dual
transformation rule of $\tse_{\rmu}^{~\hr}$.
In fact, eq.(\ref{eq:Td0'}) can be written by
\begin{equation}
 \jmath_{\hmu\hnu}=(Q^{-1})_{\hmu}^{~\rmu}\,
    \tg_{\rmu\rnu}\,({}^tQ^{-1})^{\rnu}_{~\hnu}\,,
  \quad\jmath^{\hmu\hnu}=({}^tQ)^{\hmu}_{~\rmu}\,
    \tg^{\rmu\rnu}\,(Q)_{\rnu}^{~\hnu}\,,\label{eq:gjtransf}
\end{equation}
where
\begin{eqnarray}
 \left(Q^{-1}\right)_{\hmu}^{~\rnu}&=&
    \Dpq^{-1/2}\left(\begin{array}{@{\,}cc@{\,}}
	\delta_\mu^{~\nu} & \displaystyle
	-\frac{\tg_{\mu y}+\tA_{\mu y}}{\tg_{yy}} \\[12pt]
	0 & \tg_{yy}^{-1}\end{array}\right)
   =\Dpq^{-1/2}\left(\begin{array}{@{\,}cc@{\,}}
	\delta_\mu^{~\nu} & \Dpq(\jmath_{9\mu}-\Bpq_{9\mu}) \\[10pt]
      0 & \Dpq\jmath_{99}\end{array}\right),\qquad\label{eq:Q-1}\\
 \left(Q\right)_{\rmu}^{~\hnu}&=&
    \Dpq^{1/2}\left(\begin{array}{@{\,}cc@{\,}}
	\delta_\mu^{~\nu} & \tg_{\mu y}+\tA_{\mu y} \\[10pt]
	0 & \tg_{yy}\end{array}\right)
    =\Dpq^{1/2}\left(\begin{array}{@{\,}cc@{\,}}
	\delta_\mu^{~\nu} & -\displaystyle
	\frac{\jmath_{9\mu}-\Bpq_{9\mu}}{\jmath_{99}} \\[12pt]
	0 & \Dpq^{-1}\jmath_{99}^{-1}\end{array}\right).\label{eq:Q}
\end{eqnarray}
Thus we can take the transformation rules for the vielbeins as
follows \cite{Hassan},
\begin{eqnarray}
 &&e_{\hmu}^{~\hr}=(Q^{-1})_{\hmu}^{~\rnu}\tse_{\rnu}^{~\hr}\,,
  \quad\tse_{\rmu}^{~\hr}=(Q)_{\rmu}^{~\hnu}e_{\hnu}^{~\hr}\,,\nn
 &&e_{\hr}^{~\hmu}=\tse_{\hr}^{~\rnu}(Q)_{\rnu}^{~\hmu}\,,
  \quad\tse_{\hr}^{~\rmu}=e_{\hr}^{~\hnu}(Q^{-1})_{\hnu}^{~\rmu}\,.
\label{eq:etransf}
\end{eqnarray}

We shall rewrite the 4-form and 3-form field strengths in
eq.(\ref{eq:10dFH}) and the 2-form field strength in
(\ref{eq:10dF2}).
Since we have
\begin{eqnarray}
 &&\{\thF_{\rmu\rnu\rrho\rsigma}
  +4\tA_{[\rmu}\tH_{\rnu\rrho\rsigma]}\}
    =\{\thF_{\mu\nu\rho\sigma}+4\tA_{[\mu}\tH_{\nu\rho\sigma]},\
	\thF_{\mu\nu\rho y}+3\tA_{[\mu}\tH_{\nu\rho]y}
	-\tA_{y}\tH_{\mu\nu\rho}\}\,,\nn
 &&\{\tH_{\rmu\rnu\rrho}\}\equiv\{\thF_{\rmu\rnu\rrho z}\}
     =\{\thF_{\mu\nu\rho z},\ \thF_{\mu\nu yz}\}
     =\{\tH_{\mu\nu\rho},\ \tH_{\mu\nu y}\}\,,
\end{eqnarray}
the 4-form and 3-form field strengths with the tangent indices
become
\begin{eqnarray}
 \tF_{\hr\hs\hatt\hu}&=&\tse_{\hr}^{~\rmu}\tse_{\hs}^{~\rnu}
	\tse_{\hatt}^{~\rrho}\tse_{\hu}^{~\rsigma}
	(\thF_{\rmu\rnu\rrho\rsigma}
	+4\tA_{[\rmu}\tH_{\rnu\rrho\rsigma]})\nn
  &=&-\Dpq^{-2}e_{[\hr}^{~\hmu}e_{\hs}^{~\hnu}e_{\hatt}^{~\hrho}
      e_{\hu]}^{~\hsigma}\,\ffv_{\hmu\hnu\hrho\hsigma9}
    +4\Dpq^{-1}e_{[\hr}^{~\hmu}e_{\hs}^{~\hnu}e_{\hatt}^{~\hrho}
    e_{\hu]}^{~\hsigma}\jmath_{9[\hsigma}\fth_{\hmu\hnu\hrho]}\,,\nn
 \tH_{\hr\hs\hatt}&=&\tse_{\hr}^{~\rmu}\tse_{\hs}^{~\rnu}
	\tse_{\hatt}^{~\rrho}\tH_{\rmu\rnu\rrho}\nn
  &=&\Dpq^{-3/2}\hth_{\hr\hs\hatt}
	-3\Dpq^{-1/2}\jmath_{99}^{-1}e_{[\hr}^{~\hmu}e_{\hs}^{~\hnu}
	e_{\hatt]}^{~\hrho}\jmath_{9[\hrho}
	(\Dpq^{-1}\hth_{\hmu\hnu]9}+2\p_{\hmu}\jmath_{\hnu]9})\,,
\end{eqnarray}
and similarly, the 2-form field strength of $\tA_{\hmu}$ in
eq.(\ref{eq:10dF2}) is given by
\begin{equation}
  \tF_{\hr\hs}=\tse_{\hr}^{~\rmu}\tse_{\hs}^{~\rnu}\tF_{\rmu\rnu}
    =e_{[\hr}^{~\hmu}e_{\hs]}^{~\hnu}(\Dpq^{-1}\fth_{\hmu\hnu9}
      +2\jmath_{9[\hnu}\fo_{\hmu]})\,,\label{eq:2AF2w}
\end{equation}
where
\begin{eqnarray}
 \bpq_{\hmu\hnu}&=&\hp\BNS_{\hmu\hnu}+\hq\BR_{\hmu\hnu}\,,\nn
 \hth_{\hmu\hnu\hrho}&=&3\p_{[\hmu}\bpq_{\hnu\hrho]}\,,\nn
 \ffv_{\hmu\hnu\hrho\hsigma\htau}
  &=&5\p_{[\hmu}D_{\hnu\hrho\hsigma\htau]}
    -15\,\p_{[\hmu}\bqp_{\hnu\hrho}\bpq_{\hsigma\htau]}
    +15\,\bqp_{[\hmu\hnu}\p_{\hrho}\bpq_{\hsigma\htau]}\,,\nn
 \fth_{\hmu\hnu\hrho}&=&3\p_{[\hmu}\bqp_{\hnu\hrho]}
	-3\tl\p_{[\hmu}\bpq_{\hnu\hrho]}\,,\nn
 \fo_{\hmu}&=&\p_{\hmu}\tl\,.\label{eq:2bFs}
\end{eqnarray}
Note that $\ffv$ is selfdual ($\ffv=*\ffv$)
\begin{equation}
 \ffv_{\hmu_1\hmu_2\hmu_3\hmu_4\hmu_5}
  =\frac{1}{5!}\,\frac{\epn^{\hnu_1\hnu_2\cdots\hnu_{10}}}%
	{\sqrt{-\det(\jmath_{\hmu\hnu})}}\,
    \ffv_{\hnu_1\hnu_2\hnu_3\hnu_4\hnu_5}\,
    \jmath_{\hnu_6\hmu_1}\jmath_{\hnu_7\hmu_2}\jmath_{\hnu_8\hmu_3}
	\jmath_{\hnu_9\hmu_4}\jmath_{\hnu_{10}\hmu_5}\,.
\end{equation}
Consequently, we have
\begin{eqnarray}
 \Gamma^{\hr_1\hr_2\hr_3\hr_4\hr_5}\ffv_{\hr_1\hr_2\hr_3\hr_4\hr_5}
  &=&10\ffv_{r_1r_2r_3r_4\T{9}}
	\Gamma^{r_1r_2r_3r_4\T{9}}\Gamma_+\,,\nn
 \GB^{\hmu_1\hmu_2\hmu_3\hmu_4\hmu_5}
	\ffv_{\hmu_1\hmu_2\hmu_3\hmu_49}\,\jmath_{9\hmu_5}
  &=&\Bigl(\Gamma^{r_1r_2r_3r_4\T{9}}\jmath_{99}
	+8e_{9}^{~r_4}e_{9s}\Gamma^{r_1r_2r_3s\T{9}}\Gamma_-
	+2e_{9}^{~\T{9}}e_{9r_5}\Gamma^{r_1r_2r_3r_4r_5}\Gamma_-\nn
  &&{}\quad-2e_{9}^{~s}e_{9s}\Gamma^{r_1r_2r_3r_4\T{9}}\Gamma_-\Bigr)
	\ffv_{r_1r_2r_3r_4\T{9}}\,,\label{eq:ffv}
\end{eqnarray}
where the index $\T{9}$ stands for the ninth-direction of the tangent
space.
In the above we have used the relations below,\footnote{We have
assumed that the backgrounds are independent of both $X^y$ and $\XB^9$
in (\ref{eq:d-Macr})}
\begin{eqnarray}
 \thF_{\mu\nu\rho\sigma}&=&4\p_{[\mu}\thA_{\nu\rho\sigma]}\,,\nn
 \thA_{\mu\nu\rho}&=&D_{9\mu\nu\rho}
	+\frac{3}{2}\,\Bigl\{\Bigl(\bpq_{9[\mu}\,\bqp_{\nu\rho]}
	+\frac{2\,\bpq_{9[\mu}\,\bqp_{|9|\nu}\jmath_{\rho]9}}%
	{\jmath_{99}}\Bigr)-(\bpq\leftrightarrow\bqp)\Bigr\}\,,\nn
 \thF_{\mu\nu\rho y}&=&4\p_{[\mu}\thA_{\nu\rho y]}
	=3\p_{[\mu}\thA_{\nu\rho]y}\,,\quad
 \thA_{\mu\nu y}=\bqp_{\mu\nu}
	+\frac{2\bqp_{9[\mu}\jmath_{\nu]9}}{\jmath_{99}}\nn
 \tH_{\mu\nu\rho}&=&\thF_{\mu\nu\rho z}
    =3\p_{[\mu}\thA_{\nu\rho]z}\,,\quad
 \thA_{\mu\nu z}=\bpq_{\mu\nu}
	+\frac{2\bpq_{9[\mu}\jmath_{\nu]9}}{\jmath_{99}}\,,\nn
 \tH_{\mu\nu y}&=&\thF_{\mu\nu yz}
    =2\p_{[\mu}\thA_{\nu] yz}\,,\quad
 \thA_{\mu yz}=-\frac{\jmath_{9\mu}}{\jmath_{99}}\,,\nn
 \tA_\mu&=&-\bqp_{9\mu}+\tl\bpq_{9\mu}\,,\nn
 \tA_y&=&\tl~(=\tG_{yz}/\tG_{zz})
	=\Dpq^{-2}\{(\hp l-\hq)(\hq l+\hp)+\hp\hq e^{-2\varphi}\}\,.
\end{eqnarray}
Note that in the case of $(p,q)=(1,0)$, $\bpq_{\hmu \hnu}$ is
reduced to the ordinary type IIB NSNS 2-from $\BNS_{\hmu\hnu}$ and
$\Dpq=\Delta_{(10)}=1$.

The spin connection $\tomg_{\hr}^{~\hs\hatt}$ is rewritten by
\begin{eqnarray}
 \tomg_{\hs}^{~\hu\hv}&=&\Dpq^{-1/2}\bomg_{\hs}^{~\hu\hv}
      +\frac{1}{2}\,\Dpq^{-1/2}\jmath_{99}^{-1}
	(e_{9\hs}e^{[\hu\hmu}e^{\hv]\hnu}
	+2e_{\hs}^{~\hnu}e^{[\hu\hmu}e_{9}^{~\hv]})
	(2\p_{[\hmu}\jmath_{\hnu]9}+\Dpq^{-1}\hth_{\hmu\hnu9})\nn
  &&{}-\Dpq^{-3/2}e^{[\hu\hnu}(\eta^{\hv]}_{\hs}
	-2\jmath_{99}^{-1}e_{9\hs}e_9^{\hv]})\p_{\hnu}\Dpq\,.
	\label{eq:10dIIAsc}
\end{eqnarray}
The gamma matrices with the target space indices of the dual space
are defined by
\begin{equation}
 \GB^{\hmu}=\Gamma^{\hr}e_{\hr}^{~\hmu}\,,\quad
 \GB_{\hmu}=e_{\hmu}^{~\hr}\Gamma_{\hr}=\jmath_{\hmu\hnu}\GB^{\hnu}\,,
\end{equation}
and eq.(\ref{eq:etransf}) leads to
\begin{equation}
 \tGamma^\mu=\Gamma^{\hr}\tse_{\hr}^{~\mu}
  =\Dpq^{-1/2}\,\Gamma^{\hr}e_{\hr}^{~\mu}=\Dpq^{-1/2}\GB^\mu.
\end{equation}
Similarly to the rescaling in the course of the double dimensional
reduction in the previous section, the spinors $\theta_{\pm}$ are
rescaled by $\Dpq^{1/4}$ to maintain the canonical form of the
supersymmetry transformation under T-dual.
In fact, we shall define the spinors $\btheta_{1,2}$ in type IIB
superstring theory as
\begin{equation}
 \btheta_1=\Dpq^{-1/4}\Op\theta_+\,,\quad
  \btheta_2=\Dpq^{-1/4}\theta_-\,,\label{eq:spra2b}
\end{equation}
where
\begin{equation}
  \Op=\frac{1}{\sqrt{\jmath_{99}}}\,\Gamma^{10}\GB_9\,,
  \quad\left(\Op^2=-1\right)\label{eq:Op0}
\end{equation}
and hence
\begin{equation}
 \bbtheta_1=-\bar{\theta}_+\Op\Dpq^{-1/4}\,,\quad
 \bbtheta_2=\bar{\theta}_-\Dpq^{-1/4}\,.
\end{equation}
Note that $\theta_{1,2}$ satisfy
\begin{equation}
 \Gamma_-\btheta_\xi=\theta_\xi\,,\quad
 \bbtheta_\xi\Gamma_+=\bbtheta_\xi\,,\quad
 \Gamma_+\btheta_\xi=\bbtheta_\xi\Gamma_-=0\,.\quad (\xi=1,2)
\end{equation}
One comment is in order: The $\Op$ (\ref{eq:Op0}) can be written by
\begin{equation}
   \Op=\frac{1}{\sqrt{\jmath_{99}}}\,\Gamma^{10}\GB_9
	=\frac{1}{\sqrt{\tg_{yy}}}\,\Gamma^{10}\Gamma_y\,,\label{eq:Op}
\end{equation}
which is because eq.(\ref{eq:etransf}) leads to
$\jmath_{99}^{-1/2}e_9^{~\hr}=\tg_{yy}^{-1/2}\tse_y^{~\hr}$.

Now that we shall give $\ydQ,\ydP,\ydcalG,\ydcalB$ in (\ref{eq:Td2})
by the type IIB variables. With the tangent space variables,
\begin{equation}
 \ydQ_{\hmu \hnu}=e_{\hmu}^{~\hr}e_{\hnu}^{~\hs}\ydQ_{\hr\hs}\,,\quad
 \ydP_{\hmu\hnu}=e_{\hmu}^{~\hr}e_{\hnu}^{~\hs}\ydP_{\hr\hs}\,,\quad
 \ydcalG_{j\hmu}=e_{\hmu}^{~\hr}\ydcalG_{j\hr}\,,\quad
 \ydcalB_{j\hmu}=e_{\hmu}^{~\hr}\ydcalB_{j\hr}\,,\label{eq:IIBQP}
\end{equation}
eq.(\ref{eq:Td2}) becomes
\begin{eqnarray}
 \ydQ_{\hr\hs}&=&\Dpq\Bigl\{Q_{\hr\hs}+2\jmath_{99}^{-1}
	e_9^{~\hatt}e_{9(\hr}(P_{\hs)\hatt}-Q_{\hs)\hatt})
	\Bigr\}\,,\nn
 \ydP_{\hr\hs}&=&\Dpq\Bigl\{P_{\hr\hs}+2\jmath_{99}^{-1}
	e_9^{~\hatt}e_{9[\hr}(P_{\hs]\hatt}-Q_{\hs]\hatt})
	\Bigr\}\,,\nn
 \ydcalG_{j\hr}&=&\Dpq^{1/2}\Bigl\{\calG_{j\hr}
	-\jmath_{99}^{-1}e_{9\hr}e_9^{~\hs}
	(\calG_{j\hs}+\calB_{j\hs})\Bigr\}\,,\nn
 \ydcalB_{j\hr}&=&\Dpq^{1/2}\Bigl\{\calB_{j\hr}
	-\jmath_{99}^{-1}e_{9\hr}e_9^{~\hs}
	(\calB_{j\hs}+\calG_{j\hs})\Bigr\}\,.\label{eq:2bQPGB}
\end{eqnarray}
Then we have
\begin{eqnarray}
 \ydQ_{\hr\hs}&=&Q^{(1)}_{\hr\hs}-Q^{(3)}_{\hr\hs}-Q^{(5)}_{\hr\hs}
      +\frac{i}{4}\,s^{\xi\zeta}\bbtheta_\xi
	\Gamma_{(\hr}^{\,~\hu\hv}\btheta_\zeta\hth_{\hs)\hu\hv}\nn
  &&{}+\frac{i\Dpq}{2}\,\delta^{\xi\zeta}\bbtheta_\xi\Gamma_{(\hr}
	\Gamma_{|\hatt\hu|}\btheta_\zeta \omega_{\hs)}^{\,~\hatt\hu}
	+2i\Dpq\jmath_{99}^{-1}\bbtheta_1
	\Gamma_{\hu\hv(\hr}\btheta_1e_{\hs)}^{~\hmu}e_9^{~\hv}
	\p_{\hmu}e_9^{~\hu}\,,\label{eq:Q2b}\\
 \ydP_{\hr\hs}&=&P^{(1)}_{\hr\hs}-P^{(3)}_{\hr\hs}-P^{(5)}_{\hr\hs}
       -\frac{i}{4}\,\delta^{\xi\zeta}\bbtheta_\xi
       \Gamma_{[\hr}^{\,~\hu\hv}\btheta_\zeta\hth_{\hs]\hu\hv}
    +\frac{i\Dpq}{2}\,s^{\xi\zeta}\bbtheta_\xi\Gamma_{[\hr}
       \Gamma_{|\hu\hv|}\btheta_\zeta\omega_{\hs]}^{\,~\hu\hv}\nn
  &&{}+\frac{i}{2}\,s^{\xi\zeta}\bbtheta_\xi
	\Gamma_{\hr\hs\hu}\btheta_\zeta e^{\hu\hnu}\p_{\hnu}\Dpq
      +2i\Dpq\jmath_{99}^{-1}\bbtheta_1
	\Gamma_{\hu\hv[\hr}\btheta_1e_{\hs]}^{~\hmu}e_9^{~\hv}
	\p_{\hmu}e_9^{~\hu}\,,\label{eq:P2b}\\
 \ydcalG_{j\hr}&=&i\Dpq\delta^{\xi\zeta}\bbtheta_\xi
	\Gamma_{\hr}\p_j\btheta_\zeta
    -i\Dpq\p_j\XB^{\hmu}\jmath_{99}^{-1}e_9^{~\hv}
	\p_{\hmu}e_9^{~\hu}\bbtheta_1\Gamma_{\hr\hu\hv}\btheta_1\,,
	\label{eq:G2b}\\
 \ydcalB_{j\hr}&=&i\Dpq s^{\xi\zeta}\bbtheta_\xi\Gamma_{\hr}
	\p_j\btheta_\zeta
    -i\Dpq\p_j\XB^{\hmu}\jmath_{99}^{-1}e_9^{~\hv}
	\p_{\hmu}e_9^{~\hu}\bbtheta_1\Gamma_{\hr\hu\hv}\btheta_1\,.
\label{eq:B2b}
\end{eqnarray}
where $\xi,\zeta=1,2$ and $s^{11}=-s^{22}=1,s^{12}=s^{21}=0$
(cf. (\ref{eq:2bFs})),\footnote{The extra factor $1/2$ for $n=5$ is
due to the self-duality.}
\begin{eqnarray}
  Q^{(n)}_{\hr\hs}&=&\left\{\begin{array}{ll}\displaystyle
      -\frac{i}{2\cdot n!}\,e^{\varphi(n)}\bbtheta_1
        \Gamma_{(\hr}\Gamma^{\hr_1\cdots\hr_n}\Gamma_{s)}\btheta_2
	  \,f^{(n)}_{\hr_1\cdots\hr_n}\,,&(n=1,3)\\[10pt]\displaystyle
      -\frac{i}{4\cdot n!}\,e^{\varphi(n)}\bbtheta_1
        \Gamma_{(\hr}\Gamma^{\hr_1\cdots\hr_n}\Gamma_{s)}\btheta_2
	\,f^{(n)}_{\hr_1\cdots\hr_n}\,,&(n=5)\end{array}
	\right.\nn[10pt]
  P^{(n)}_{\hr\hs}&=&\left\{\begin{array}{ll}\displaystyle
      -\frac{i}{2\cdot n!}\,e^{\varphi(n)}\,\bbtheta_1
	\Gamma_{[\hr}\Gamma^{\hr_1\cdots\hr_n}\Gamma_{s]}\btheta_2
	 \,f^{(n)}_{\hr_1\cdots\hr_n}\,,&(n=1,3)\\[10pt]\displaystyle
      -\frac{i}{4\cdot n!}\,e^{\varphi(n)}\,\bbtheta_1
	\Gamma_{[\hr}\Gamma^{\hr_1\cdots\hr_n}\Gamma_{s]}\btheta_2
	 \,f^{(n)}_{\hr_1\cdots\hr_n}\,,&(n=5)\end{array}\right.
\end{eqnarray}
$\varphi(n)$ is a ``coupling parameter'' given by
\begin{equation}
 \varphi(n)=\varphi+\frac{7-n}{2}\,\log\Dpq\,,\label{eq:IIBcp}
\end{equation}
and $\omega_{\hr}^{~\hs\hatt}$ is the spin connection in type IIB
theory in ten dimensions.
Note that $\varphi(n)$ and $h_{\hr\hs\hatt}$ are reduced to the type
IIB dilaton and the field strength of the NSNS 2-form field,
respectively, in the case of $(p,q)=(1,0)$, or the fundamental string.

Then, the $\theta^2$-order part of (\ref{eq:ddrac}) is given by
\begin{eqnarray}
 S_{ddr}\big|_{\theta^2}&=&-2\pi iT\int\!d\sigma^0\!\int_0^{2\pi}
    \hspace{-1ex}d\sigma^1\,\zl\Biggl[\Dpq(\sqrt{-\tgamma}\,
    \tgamma^{ij}\delta^{\xi\zeta}+\epn^{ij}s^{\xi\zeta})
    \p_i\XB^{\hmu}\bbtheta_\xi\GB_{\hmu}\DB_j \btheta_\zeta\nn
  &&{}+\frac{1}{2}\,
    \Biggl\{\frac{1}{4}\,(\sqrt{-\tgamma}\,
    \tgamma^{ij}s^{\xi\zeta}+\epn^{ij}\delta^{\xi\zeta})
	\p_i\XB^{\hmu}\p_j\XB^{\hnu}
    \bbtheta_\xi\GB_{\hmu}^{~\hrho\hsigma}\btheta_\zeta
	\hth_{\hnu\hrho\hsigma}\nn
 &&{}\quad+(\sqrt{-\tgamma}\,\tgamma^{ij}+\epn^{ij})\,
    \p_i\XB^{\hmu}\p_j\XB^{\hnu}\Bigl(
	\frac{e^{\varphi(1)}}{2}\,\bbtheta_1\GB_{\hmu}
	\GB^{\hmu_1}\GB_{\hnu}\btheta_2 \fo_{\hmu_1}\nn
  &&{}\qquad-\frac{e^{\varphi(3)}}{2\cdot 3!}\,\bbtheta_1\GB_{\hmu}
	\GB^{\hmu_1\hmu_2\hmu_3}\GB_{\hnu}\btheta_2
	\fth_{\hmu_1\hmu_2\hmu_3}
      -\frac{e^{\varphi(5)}}{4\cdot 5!}\,\bbtheta_1\GB_{\hmu}
	\GB^{\hmu_1\cdots\hmu_5}\GB_{\hnu}\btheta_2
	\ffv_{\hmu_1\cdots\hmu_5}\Bigr)\nn
 &&{}\quad+\frac{1}{2}\,\epn^{ij}s^{\xi\zeta}\p_i\XB^{\hmu}
    \p_j\XB^{\hnu}\bbtheta_\xi\GB_{\hmu\hnu}{}^{\hrho}\btheta_\zeta
	\p_{\hrho}\Dpq\Biggr\}\Biggr]\,,\label{eq:pqac2}
\end{eqnarray}
where
\begin{eqnarray}
 \DB_i=\p_i+\frac{1}{4}\,\p_i\XB^{\hmu}\,
	\omega_{\hmu}^{~\hr\hs}\Gamma_{\hr\hs}\,.
\end{eqnarray}
As is noted in Ref.\cite{CLPS}, the derivative for chiral spinor
appears only in the covariant derivative.
Note that in the case of $(p,q)=(1,0)$, or fundamental string, the
resulting action (\ref{eq:pqac2}) is reduced to the one in
Ref.\cite{CLPS}.
In that case we can see that the fundamental string couples with the
RR 1-, 3- and  5-form fields with strength $e^{\varphi}$.
In the case of $(p,q)=(0,1)$, or D-string, the strengths of the
coupling with the RR 1-, 3- and 5-form field strengths are given by
$e^{-2\varphi}$, $e^{-\varphi}$, $1$, respectively.
Putting (\ref{eq:pqac0}) and (\ref{eq:pqac2}) together, we have
explicitly the Green-Schwarz type $(p,q)$-string action of order up to
quadratic in $\theta$ in type IIB superstring
\begin{eqnarray}
 S_{ddr}&=&-\frac{2\pi T}{2}\int\!d\sigma^0\!\int_0^{2\pi}
    \hspace{-1ex}d\sigma^1\,{\zl}\,\Dpq\Biggl[\sqrt{-\tgamma}\,
      \tgamma^{ij}\,\p_i\XB^{\hmu}\p_j\XB^{\hnu}\jmath_{\hmu\hnu}
      -\epn^{ij}\p_i\XB^{\hmu}\p_j\XB^{\hnu}\Bpq_{\hmu\hnu}\nn
  &&{}\quad+2i(\sqrt{-\tgamma}\,
    \tgamma^{ij}\delta^{\xi\zeta}+\epn^{ij}s^{\xi\zeta})
    \p_i\XB^{\hmu}\bbtheta_\xi\GB_{\hmu}\DB_j \btheta_\zeta\nn
 &&{}\quad+\frac{i}{\Dpq}\,
    \Bigl\{\frac{1}{4}\,(\sqrt{-\tgamma}\,
    \tgamma^{ij}s^{\xi\zeta}+\epn^{ij}\delta^{\xi\zeta})
	\p_i\XB^{\hmu}\p_j\XB^{\hnu}
    \bbtheta_\xi\GB_{\hmu}^{~\hrho\hsigma}\btheta_\zeta
	\hth_{\hnu\hrho\hsigma}\nn
 &&{}\qquad+(\sqrt{-\tgamma}\,\tgamma^{ij}+\epn^{ij})\,
    \p_i\XB^{\hmu}\p_j\XB^{\hnu}\Bigl(
	\frac{e^{\varphi(1)}}{2}\,\bbtheta_1\GB_{\hmu}
	\GB^{\hmu_1}\GB_{\hnu}\btheta_2\fo_{\hmu_1}\nn
  &&{}\quad\qquad-\frac{e^{\varphi(3)}}{2\cdot 3!}\,
	\bbtheta_1\GB_{\hmu}\GB^{\hmu_1\hmu_2\hmu_3}\GB_{\hnu}
	\btheta_2\fth_{\hmu_1\hmu_2\hmu_3}
      -\frac{e^{\varphi(5)}}{4\cdot 5!}\,\bbtheta_1\GB_{\hmu}
	\GB^{\hmu_1\cdots\hmu_5}\GB_{\hnu}\btheta_2
	\ffv_{\hmu_1\cdots\hmu_5}\Bigr)\nn
 &&{}\qquad+\frac{1}{2}\,\epn^{ij}s^{\xi\zeta}\p_i\XB^{\hmu}
    \p_j\XB^{\hnu}\bbtheta_\xi\GB_{\hmu\hnu}{}^{\hrho}\btheta_\zeta
	\p_{\hrho}\Dpq\Bigr\}\Biggr]\,.\label{eq:pqacP}
\end{eqnarray}

Finally in this subsection, we also give the action (\ref{eq:ddrac})
in Nambu-Goto form up to quadratic in $\theta$ by integrating out the
worldsheet metric $\tgamma_{ij}$ in (\ref{eq:ddrac}), or in
(\ref{eq:pqacP}) (cf. Appendix \ref{S:Cal2b})
\begin{eqnarray}
 S_{ddr}=2\pi T\!\int\!d^2\sigma\,{\zl}\,\Dpq\Bigl[
	-\sqrt{-\jmath}+\frac{1}{2}\,\epn^{ij}\Bpq_{ij}
    -2i\sqrt{-\jmath}\jmath^{ij}\bbtheta\PB\GB_i\nB_j\btheta\Bigr]\,,
 \label{eq:pqacNG}
\end{eqnarray}
where
\begin{eqnarray}
 \PB&=&\frac{1}{2}\,\Bigl(1+\frac{1}{2\sqrt{-\jmath}}\,\sigma_3
	\otimes\epn^{ij}\GB_{ij}\Gamma^{10}\Bigr),\\
 \DB_i&=&\p_i+\frac{1}{4}\,\p_i\XB^{\hmu}\,
	\omega_{\hmu}^{~\hr\hs}\Gamma_{\hr\hs}\,,\\
 \nB_i&=&\DB_i-\frac{1}{4}\,\p_{\hmu}(\ln\Dpq)\GB^{\hmu}\GB_i
	+\frac{1}{8\Dpq}\,
	\sigma_3\otimes\GB^{\hrho\hsigma}\hth_{i\hrho\hsigma}\nn
 &&{}+\frac{e^{\varphi}}{8}\Bigl\{
	i\sigma_2\otimes\Dpq^2\,\GB^{\hmu_1}\fo_{\hmu_1}
      -\sigma_1\otimes\frac{\Dpq}{3!}\,
	\GB^{\hmu_1\hmu_2\hmu_3}\fth_{\hmu_1\hmu_2\hmu_3}\nn
  &&{}\qquad-i\sigma_2\otimes\frac{1}{2\cdot 5!}\,
    \GB^{\hmu_1\cdots\hmu_5}\ffv_{\hmu_1\cdots\hmu_5}\Bigr\}\,
	\GB_i\,,\\
 \jmath&=&\det\jmath_{ij}\,,\quad\GB_i=\p_i\XB^{\hmu}\,\GB_{\hmu},\\
 \jmath_{ij}&=&\p_i\XB^{\hmu}\p_j\XB^{\hnu}\jmath_{\hmu\hnu}\,,
	\quad\Bpq_{ij}=\p_i\XB^{\hmu}\p_j\XB^{\hnu}\Bpq_{\hmu\hnu},
 \quad\hth_{i\hnu\hrho}=\p_i\XB^{\hmu}\hth_{\hmu\hnu\hrho}\,,
\end{eqnarray}
$\sigma_{\{1,2,3\}}$ are Pauli matrices and
\begin{equation}
 \quad\bbtheta=\left(\bbtheta_1~\,\bbtheta_2\right)\,,
  \quad\btheta=\left(\begin{array}{@{\,}c@{\,}}
	\btheta_1\\ \btheta_2 \end{array}\right).\label{eq:sp2b}
\end{equation}

\subsection{Fermionic symmetry}
First, we show that the ($p,q$)-string action (\ref{eq:pqacNG}) has
the $\kappa$-symmetry which is really inherited from the
$\kappa$-symmetry in the supermembrane (\ref{eq:K11}).
In fact, we shall analyse the $\kappa$-symmetry of (\ref{eq:d-Macr})
along the argument in \cite{BT}.
Since (\ref{eq:d-Macr}) is equivalent to (\ref{eq:Macr2}) when
$\epn^{ij}\p_i\Xi_j=0$ and (\ref{eq:Macr2}) is invariant under the
$\kappa$-transformation (\ref{eq:Akappa}), the variation of
(\ref{eq:d-Macr}) under the $\kappa$-transformation should be
proportional to $\epn^{ij}\p_i\Xi_j$\,,
\begin{eqnarray}
 \dkp S_{ddr}&=& -2\pi T\!\int\!d^2\sigma\,\zl\,\epn^{ij}\p_i\Xi_j\,
	(\dkp X^\mu\tA_{\mu y}
	+i\bar{\theta}\Gamma_y\Gamma^{10}\dkp\theta-\dkp\XB^9)\nn
  &=&2\pi T\!\int\!d^2\sigma\,\zl\,\epn^{ij}\p_i\Xi_j\{
	i\bar{\theta}(\Gamma^{\mu}\tA_{\mu y}
	-\Gamma_y\Gamma^{10})(1+\Gamma_F)\kappa+\dkp\XB^9\}\,,
\end{eqnarray}
where
\begin{eqnarray}
  \theta=\theta_++\theta_-\,,\quad \kappa=\kappa_++\kappa_-\,.
\end{eqnarray}
Thus, $\XB^9$ should be transformed as
\begin{equation}
 \dkp\XB^9=-i\bar{\theta}(\Gamma^{\mu}\tA_{\mu y}
	-\Gamma_y\Gamma^{10})(1+\Gamma_F)\kappa
  \equiv-i\bar{\theta} M(1+\Gamma_F)\kappa\,.\label{eq:dkX90}
\end{equation}
We shall rewrite eq.(\ref{eq:dkX90}) with the dual variables, or the
($p,q$)-string fields and the supergravity background fields.
Since the spinors are converted as in eq.(\ref{eq:spra2b}), we
shall write in the matrix form (\ref{eq:sp2b}).
Then, we have\footnote{In rewriting the relations with the dual
variables we have used the relation (cf. (\ref{eq:d-Y}))
\begin{equation}
 \p_iX^y\sim\frac{\tgamma_{ij}\epn^{kj}}{\tbG_{yy}\sqrt{-\tgamma}}\,
	(\p_k X^\mu\tbB_{\mu y}-\p_k\XB^9)
    -\frac{\tbG_{\mu y}}{\tbG_{yy}}\,\p_i X^\mu\,.
\end{equation}} (cf. Appendix \ref{S:Calkp})
\begin{eqnarray}
 \dkp\XB^9 &=&-i\Dpq^{1/2}\,\bbtheta\left(\begin{array}{@{\,}cc@{\,}}
	-\Op M(1+\Gamma_F)\Op & -\Op M(1+\Gamma_F) \\[5pt]
	M(1+\Gamma_F)\Op & M(1+\Gamma_F)
	\end{array}\right)\bkappa\nn
 &=&-i\bbtheta\,\GB^9\left(\begin{array}{@{\,}cc@{\,}}
	1-\Op\Gamma_F\Op & 0 \\[5pt]
	0 & 1+\Gamma_F\end{array}\right)\bkappa\nn
 &=&-i \bbtheta\GB^9 \left(\begin{array}{@{\,}cc@{\,}}
	1-\Gamma_B & 0 \\[5pt]
	0 & 1+\Gamma_B \end{array}\right)\bkappa'
  \equiv -i\bbtheta\GB^9(1-\Gamma_{\mathrm{IIB}})\bkappa'\,,
 \label{eq:kpX9}
\end{eqnarray}
where
\begin{eqnarray}
 &&\btheta =\left(\begin{array}{@{\,}c@{\,}} \btheta_1\\[5pt]
	\btheta_2 \end{array}\right),\quad
 \bkappa=\left(\begin{array}{@{\,}c@{\,}}
	\bkappa_1 \\[5pt] \bkappa_2 \end{array}\right),\quad
 \bkappa'=\frac{\Gamma_-}{(1-\jmath_{99}^{-1}\jmath^{kl}
	\jmath_{k9}\jmath_{l9})}\left(\begin{array}{@{\,}c@{\,}}
	\Bigl(1-\frac{\epn^{mn}\jmath_{n9}\GB_m\Op}%
    {\sqrt{-\jmath}\sqrt{\jmath_{99}}}\Bigr)\kappa_1 \\[8pt]
	\Bigl(1+\frac{\epn^{mn}\jmath_{n9}\GB_m\Op}%
    {\sqrt{-\jmath}\sqrt{\jmath_{99}}}\Bigr)\kappa_2
	\end{array}\right),\qquad\label{eq:bkappa'}\\
 &&\Gamma_B=\frac{1}{2}\,\frac{1}{\sqrt{-\jmath}}\,
	\epn^{ij}\GB_{ij}\Gamma^{10}\,,
    \quad\Gamma_{\mathrm{IIB}}=\sigma_3\otimes\Gamma_B\,,
\end{eqnarray}
and we have used the following relations,
\begin{eqnarray}
 M&\hspace{-1.5ex}\Bigl(=&\Gamma^{\mu}\tA_{\mu y}
	-\Gamma_y\Gamma^{10}\Bigl)\,
    =\Dpq^{-1/2}(\GB^9+2\jmath_{99}^{-1/2}\Op\Gamma_+)\,,\\
 \Op\GB^9&=&\GB^9\Op+2\jmath_{99}^{-1/2}\Gamma^{10}\,,\\
 (1+\Gamma_F)\,\Gamma_-&=&(1+\Gamma_B)\,\frac{1+\frac{\epn^{ij}
    \jmath_{j9}\GB_i\Op}{\sqrt{-\jmath}\sqrt{\jmath_{99}}}}%
      {1-\jmath^{kl}\jmath_{99}^{-1}\jmath_{k9}\jmath_{l9}}\,
	\Gamma_-\,,\\
 (1-\Op\Gamma_F\Op)\,\Gamma_-&=&(1-\Gamma_B)\,
	\frac{1-\frac{\epn^{mn}\jmath_{n9}\GB_m\Op}%
    {\sqrt{-\jmath}\sqrt{\jmath_{99}}}}{1-\jmath^{kl}\jmath_{99}^{-1}
	\jmath_{k9}\jmath_{l9}}\,\Gamma_-\,.
\end{eqnarray}
We also calculate $\dkp X^{\mu}\,(=\dkp\XB^{\mu})$ to get
\begin{eqnarray}
 \dkp X^{\mu}&=&-i(\bar{\theta}_++\bar{\theta}_-)
	\Gamma^{\mu}(1+\Gamma_{F})(\kappa_++\kappa_-)\nn
  &=&-i\Dpq^{1/2}\left(\bbtheta_1~\,\bbtheta_2\right)
    \left(\begin{array}{@{\,}cc@{\,}}
	-\Op\Gamma^\mu(1+\Gamma_F)\Op & \Op\Gamma^\mu
	(1+\Gamma_F)\\[5pt]-\Gamma^\mu(1+\Gamma_F)\Op &
	\Gamma^\mu(1+\Gamma_F)\end{array}\right)
    \left(\begin{array}{@{\,}c@{\,}}
	\bkappa_1 \\[5pt] \bkappa_2 \end{array}\right)\,\nn
  &=&-i\bbtheta\,\GB^\mu(1-\Gamma_{\mathrm{IIB}})\bkappa'\,,
\label{eq:kappaX}
\end{eqnarray}
where the following relations have been used,
\begin{equation}
 \Op\Gamma^{\mu}=\Gamma^{\mu}\Op\,,\quad
	\Gamma^\mu=\Dpq^{-1/2}\GB^{\mu}\,.
\end{equation}
The $\kappa$-transformations of the fermionic coordinates are
given by
\begin{eqnarray}
 \dkp\btheta_1&=&\dkp(\Dpq^{-1/4}\Op\theta_+)
	=(1-\Op\Gamma_F\Op)\kappa_1+O(\theta^2)\,,\nn
 \dkp\btheta_2&=&\dkp(\Dpq^{-1/4}\theta_-)
	=(1+\Gamma_F)\bkappa_2-\frac{1}{4}\,\dkp(\ln\Dpq)\,\btheta_2
	=(1+\Gamma_F)\bkappa_2+O(\theta^2),\quad
\end{eqnarray}
and hence we arrive at the conclusion that the ($p,q$)-string action
(\ref{eq:pqacNG}) is invariant under the $\kappa$-transformation,
which is really inherited from that of the supermembrane
\begin{eqnarray}
 \dkp\btheta = (1-\Gamma_{\mathrm{IIB}})\bkappa'\,,
  \quad\dkp X^{\hmu}=-i\bbtheta\Gamma^{\hmu}
	(1-\Gamma_{\mathrm{IIB}})\bkappa'\,,\quad
 \dkp\Phi_{bg}=\dkp X^{\hmu}\p_{\hmu}\Phi_{bg}\,.
\end{eqnarray}

Next, we consider the SUSY transformation.
Similarly, the variation of (\ref{eq:d-Macr}) under the SUSY
transformation should be proportional to $\epn^{ij}\p_i\Xi_j$ and it
is calculated as (cf. eq.(\ref{eq:2aSS}))
\begin{eqnarray}
 \dss S_{ddr}&=&-2\pi T\!\int\!d^2\sigma\,\zl\,\epn^{ij}\p_i\Xi_j\,
	(\dss X^\mu\tA_{\mu y}
    +i\bar{\theta}\Gamma_y\Gamma^{10}\dss\theta-\dss\XB^9)\nn
  &=&2\pi T\!\int\!d^2\sigma\,\zl\,\epn^{ij}\p_i\Xi_j\,
	\{i\bar{\theta}(\Gamma^\mu\tA_{\mu y}
    -\Gamma_y\Gamma^{10})\epn+\dss\XB^9\}\,,
\end{eqnarray}
and hence
\begin{eqnarray}
 \dss\XB^9 &=&-i(\bar{\theta}_++\bar{\theta}_-)M(\epn_++\epn_-)\nn
  &=&-i\left(\bbtheta_1~\,\bbtheta_2\right)
    \left(\begin{array}{@{\,}cc@{\,}}
	-\Op\{\GB^9+2\jmath_{99}^{-1/2}\Op\Gamma_+\}\Op & 0\\[5pt]
	0 & \GB^9+2\jmath_{99}^{-1/2}\Op\Gamma_+
	\end{array}\right)\left(\begin{array}{@{\,}c@{\,}}
	\bepn_1 \\[5pt] \bepn_2 \end{array}\right)\,\nn
  &=&-i\bbtheta\GB^9\bepn=i\bbepn\GB^9\btheta\,.
\end{eqnarray}
We also have
\begin{equation}
 \dss X^{\mu}=i(\bar{\epn}_++\bar{\epn}_-)\Gamma^{\mu}
	(\theta_++\theta_-)
    =i\left(\bbepn_1~\,\bbepn_2\right)
	\GB^{\mu}\left(\begin{array}{@{\,}c@{\,}}
	\btheta_1 \\[5pt] \btheta_2 \end{array}\right).
\end{equation}
Thus, the ($p,q$)-string action (\ref{eq:pqacNG}) is invariant under
the following SUSY transformation, which is inherited from the
superdiffeomorphism in the supermembrane action,
\begin{eqnarray}
 \dss\btheta = \bepn\,,\quad
 \dss\XB^{\hmu} = i\bbepn\Gamma^{\hmu}\btheta\,,\quad
 \dss\Phi_{bg} = \dss\XB^{\hmu}\p_{\hmu}\Phi_{bg}\,.
\end{eqnarray}

\section{Summary and discussion}
In this paper we have explicitly derived the ($p,q$)-string action of
the Green-Schwarz type from the supermembrane action up to quadratic
order in the anti-commuting supercoordinate in the bosonic curved
background.
We have also shown that both the $\kappa$-symmetry and the
supersymmetry in the ($p,q$)-string action are really inherited from
the $\kappa$-symmetry and the (super) diffeomorphism in the
supermembrane action, respectively.
In fact, we have first studied the double dimensional reduction of the
wrapped supermembrane compactified on a 2-torus up to quadratic order
of the anti-commuting coordinate.\footnote{The procedure of the double
dimensional reduction here was realized on the bosonic sector of the
matrix-regularized wrapped supermembrane on $\R^9\times T^2$
\cite{OUY2} relying on the technique given in \cite{UY4,Ced,UY3}.}
Next, we applied the T-dual transformation and explicitly derived the
type IIB Green-Schwarz superstring action for the $(p,q)$-string in
eq.(\ref{eq:pqacNG}).\footnote{In the case of $(p,q)=(1,0)$, or the
fundamental string, the resulting action is reduced to the one in
\cite{CLPS}.}
This indicates that the supermembrane actually includes a
$(p,q)$-string as an excitation mode or object.
The (1,0)-string (F-string) is, of course, a fundamental mode
in the weak coupling region $\gIIB\ll1$, while the (0,1)-string
(D-string) in the strong coupling region $\gIIB\gg1$ for $l=0$.
However, the valid region to treat the $(p,q)$-string perturbatively
is still obscure and it is deserved to be investigated
further.\footnote{Of course, a BPS saturated classical solution of the
$(p,q)$-string action (\ref{eq:pqacNG}) is valid irrespective of the
value of the string coupling $\gIIB$.}

In this paper we have considered classically to approach the boundary
of vanishing cycles of the 2-torus with the wrapped supermembrane.
On the other hand, Refs.\cite{SY,UY} studied quantum mechanical
justification of the double dimensional reduction in Ref.\cite{DHIS}.
In those references, the Kaluza-Klein modes associated with the
$\sigma^2$-coordinate were not removed classically, but they were
integrated in the path integral formulation of the wrapped
supermembrane theory.
Similar quantum mechanical investigation of the double dimensional
reduction adopted in this paper deserves to be investigated.

\vspace{\baselineskip}
\noindent{\bf Acknowledgments:}
This work is supported in part by MEXT Grant-in-Aid for
the Scientific Research \#20540249 (S.U.).

\appendix
\section{Notation\label{S:N}}
11d super spacetime indices:
\begin{eqnarray}
 \hM&=&(M,\alpha)\,,\nn
  M,N,P,Q &=& 0,1,\cdots,8,9,10\,, \nn
  \alpha ,\beta ,\gamma &=& 1, 2, \cdots, 32\,.
\end{eqnarray}
11d tangent superspace indices:
\begin{eqnarray}
  \hA&=&(A,a)\,,\nn
  A,B,C &=& 0,1,\cdots,8,9,10\,, \nn
  a,b,c &=& 1, 2, \cdots, 32\,.
\end{eqnarray}
11d rotated spacetime indices:
\begin{eqnarray}
  \RM,\RN,\RP,\RQ &=& 0,1,\cdots,8,y,z\,.
\end{eqnarray}
10d spacetime and tangent space indices:
\begin{eqnarray}
  \hmu,\hnu &=& 0,1,\cdots,8,9\,, \nn
  \hr,\hs &=& 0,1,\cdots,8,9\,.
\end{eqnarray}
10d rotated super spacetime indices:
\begin{eqnarray}
  \tm&=&(\rmu,\alpha)\,,\nn
  \rmu,\rnu &=& 0,1,\cdots,8,y\,.
\end{eqnarray}
9d spacetime and tangent space indices:
\begin{eqnarray}
\mu,\nu&=& 0,1,\cdots 8\,,\nn
r,s&=& 0,1,\cdots 8\,.
\end{eqnarray}
The worldvolume and worldsheet indices:
\begin{eqnarray}
  \hi,\hj,\hk  &=& 0,1,2\,, \nn
  i,j,k &=& 0,1\,.
\end{eqnarray}
target space metrics:
\begin{eqnarray}
  G&=& \textrm{11d target space metric} \,,\nn
  \tG&=& \textrm{11d rotated target space metric} \,,\nn
  g&=& \textrm{10d IIA target space metric}\,,\nn
  \tg &=& \textrm{10d IIA rotated target space metric} \,,\nn
  \jmath &=& \textrm{10d IIB target space metric} \,.
\end{eqnarray}
Worldvolume and worldsheet metrics:
\begin{eqnarray}
\hat{\gamma} &=& \textrm{membrane worldvolume metric}\,,\nn
\gamma &=& \textrm{string worldsheet metric}\,.
\end{eqnarray}
(Anti-)symmetrization r.w.t.\ indices:
\begin{eqnarray}
   A_{[\mu} B_{\nu]} &=& \frac{1}{2}\left(A_{\mu} B_{\nu}
	- A_{\nu} B_{\mu}\right)\,,\nn
   A_{[\mu} B_{\nu} C_{\rho]} &=& \frac{1}{3!}
	(A_{\mu} B_{\nu}C_{\rho}
	+A_{\nu} B_{\rho}C_{\mu}+A_{\rho} B_{\mu}C_{\nu}\nn
    &&\quad - A_{\mu}B_{\rho}C_{\nu}
	-A_{\rho} B_{\nu}C_{\mu}-A_{\nu} B_{\mu}C_{\rho})\,,\nn
   A_{[\mu} B_{|\nu|} C_{\rho]} &=& \frac{1}{2}
	(A_{\mu} B_{\nu}C_{\rho}-A_{\rho} B_{\nu}C_{\mu})\,,\nn
   A_{(\mu} B_{\nu)} &=& \frac{1}{2}\left(A_{\mu} B_{\nu}
	+ A_{\nu} B_{\mu}\right)\,,\quad\mbox{etc.}\,.
\end{eqnarray}

\section{Bispinor formula}\label{S:Bisp}
The charge conjugate matrix $C$ satisfies
\begin{equation}
 C\Gamma_{A}\,C^{-1}=-{}^t\Gamma_{A}\,,\quad {}^tC=-C\,,\label{eq:CC}
\end{equation}
or
\begin{equation}
 C_{\alpha\gamma}(\Gamma_{A})^{\gamma}_{~\tau}C^{\tau\beta}
    =-(\Gamma_{A})^\beta_{~\alpha} \,,\quad C_{\alpha\beta}
    =-C_{\beta\alpha}\,.\label{eq:CC2}
\end{equation}
Eq.(\ref{eq:CC}) leads to
\begin{equation}
 \bar{\theta}\Gamma_{A_1A_2\cdots A_n} \psi=(-1)^{\frac{n(n+1)}{2}}
 \bar{\psi} \Gamma_{A_1A_2\cdots A_n} \theta\,.\label{eq:bisp1}
\end{equation}
where $\theta$ and $\psi$ are 32-component Majorana spinor in
eleven dimensions and
\begin{equation}
 \bar{\theta}=i\,{}^t\theta\,C=i\,{}^t\theta\, \Gamma^0.\quad
	(\bar{\theta}_\alpha=i\theta^\beta C_{\beta \alpha})
\end{equation}
Note that for $n=0,3,4,7,8$ and $n=1,2,5,6,9,10$, the bispinor
products eq.(\ref{eq:bisp1}) are symmetric and antisymmetric,
respectively and
\begin{equation}
  (\Gamma_{A_1A_2\cdots A_n})_{\alpha \beta}
    =\left\{\begin{array}{l@{\quad}l}
	-(\Gamma_{A_1A_2\cdots A_n})_{\beta\alpha}\,,&
		(n=0,3,4,7,8)\\[5pt]
	(\Gamma_{A_1A_2\cdots A_n})_{\beta\alpha}\,,&
		(n=1,2,5,6,9,10)\end{array}\right.\label{eq:Gammaab}
\end{equation}
where the spinor indices are lowered and raised by $C$
\begin{eqnarray}
  M_{\alpha \beta}&=&C_{\alpha\gamma}M^\gamma{}_\beta\,,\\
  M^{\alpha \beta}&=&M^\alpha{}_\gamma C^{\gamma \beta},\\
  M_\alpha{}^\beta&=& C_{\alpha\gamma}M^\gamma{}_\delta
	C^{\delta\beta}.
\end{eqnarray}
In particular we have
\begin{equation}
  (\Gamma^A)_{\alpha\beta}=(\Gamma^A)_{\beta \alpha}\,.
\end{equation}
Putting $\theta=\psi$ in eq.(\ref{eq:bisp1}) we have the identity
\begin{equation}
  \bar{\theta}\Gamma^{A_1 A_2 \cdots A_n}\theta=0\,.\quad
  	(n=1,2,5,6,9,10)\label{eq:bisp2}
\end{equation}
For the chiral-projected spinors $\psi_{\pm}, \theta_{\pm}$
(cf. eq.(\ref{eq:chsp})), eq.(\ref{eq:CC}) leads to
\begin{eqnarray}
 \bar{\psi}_{\pm}\Gamma_{A_1A_2\cdots A_n} \theta_{\pm}&=&0\,,
     \quad (n \in 2\N) \nn
 \bar{\psi}_{\pm}\Gamma_{A_1A_2\cdots A_n} \theta_{\mp}&=&0\,,
      \quad (n \in 2\N+1) \label{eq:bisp}
\end{eqnarray}
and from eqs.(\ref{eq:bisp1}) and (\ref{eq:bisp}) we have
\begin{eqnarray}
 \bar{\theta}_{+}\Gamma_{A_1A_2\cdots A_n} \psi_{-}
  &=&\left\{\begin{array}{l@{\quad}l}
	\bar{\psi}_{-}\Gamma_{A_1A_2\cdots A_n} \theta_{+}\,,&
		(n=0,4,8)\\[5pt]
	-\bar{\psi}_{-}\Gamma_{A_1A_2\cdots A_n} \theta_{+}\,,&
		(n=2,6,10)\end{array}\right.\nn
 \bar{\theta}_{+}\Gamma_{A_1A_2\cdots A_n} \psi_{+}
  &=&\left\{\begin{array}{l@{\quad}l}
	-\bar{\psi}_{+}\Gamma_{A_1A_2\cdots A_n} \theta_{+}\,,&
		(n=1,5,9)\\[5pt]
	\bar{\psi}_{+}\Gamma_{A_1A_2\cdots A_n} \theta_{+}\,.&
		(n=3,7)\end{array}\right.
\end{eqnarray}
We also write down some useful formulas for the $\Gamma$ matrices
\begin{eqnarray}
 \Gamma_{\hr}\Gamma^{\hr_1\hr_2\cdots\hr_n}
  &=&\Gamma_{\hr}^{~\hr_1\cdots\hr_n}
  +\sum_i(-1)^{i-1}\delta_{\hr}^{\hr_i}
   \Gamma^{~\hr_1\cdots\stackrel{i}{\check{\hr_i}}\cdots\hr_n}\,,\nn
 \Gamma_{\hr\hs}\Gamma^{\hr_1\cdots\hr_n}
  &=&\Gamma_{\hr\hs}^{~~\hr_1\hr_2\cdots\hr_n}
	+2\sum_i(-1)^{i}\delta_{[\hr}^{\hr_i}
  \Gamma_{\hs]}^{~\hr_1\cdots\stackrel{i}{\check{\hr_i}}\cdots\hr_n}\nn
  &&{}+2\sum_{i<j}(-1)^{i+j}\delta_{[\hr}^{\hr_i}\delta_{\hs]}^{\hr_j}
   \Gamma^{~\hr_1\cdots\stackrel{i}{\check{\hr_i}}\cdots
	\stackrel{j}{\check{\hr_j}}\cdots\hr_n}.
\end{eqnarray}

\section{11d vs. 10d background fields}\label{S:R}
The 11-dimensional metric can be written by
\begin{eqnarray}
 \hG_{MN}&\equiv& e^{-\frac{2}{3}\phi}
    \left(\begin{array}{@{\,}cc@{\,}}
	\ga_{\hmu\hnu}+e^{2\phi}A_{\hmu}A_{\hnu}&
		 e^{2\phi}  A_{\hmu} \\[10pt]
	e^{2\phi}  A_{\hnu} & e^{2\phi} \end{array}\right)\nn
   &=& \left(\begin{array}{@{\,}cc@{\,}}
	\frac{1}{\sqrt{\hG_{1010}}}\,\ga_{\hmu\hnu}
	+\frac{1}{\hG_{1010}}\hG_{\hmu10}\hG_{\hnu10}&
		 \hG_{\hmu10} \\[10pt]
	\hG_{\hnu10} & \hG_{1010} \end{array}\right),
    \label{eq:KKmetric}
\end{eqnarray}
and the third-rank and forth-rank antisymmetric tensors $\hA_{MNP},
\hF_{MNPQ}$ are decomposed as
\begin{eqnarray}
 \{\hA_{MNP}\} &=& \{\hA_{\mu\nu\rho},
     \hA_{\mu\nu10},\hA_{\mu\nu9},\hA_{\mu910}\}
  =\{C_{\mu\nu\rho},
	B_{\mu\nu},C_{\mu\nu9},B_{\mu9}\}\,,\label{eq:dhA}\\
 \{\hF_{MNPQ}\}
  &=&\{\hF_{\hmu\hnu\hrho\hsigma}, \hF_{\hmu\hnu\hrho 10}\}
  =\{F_{\hmu\hnu\hrho\hsigma}, H_{\hmu \hnu \hrho}\}
  =\{4\p_{[\hmu}C_{\hnu\hrho\hsigma]},3\p_{\hmu}B_{\hnu\hrho]}\}\,.
\end{eqnarray}
Those fields are related to those in type IIB theory as
\begin{eqnarray}
  \ga_{\mu\nu}&=& \jmath_{\mu\nu}
    -\frac{\jmath_{9\mu}\jmath_{9\nu}
	-\BNS_{9\mu}\BNS_{9\nu}}{\jmath_{99}}\,,\\
  \ga_{9\mu}&=& \frac{\BNS_{9\mu}}{\jmath_{99}}\,,\\
  \ga_{99}&=&\frac{1}{\jmath_{99}}\,,\\
  C_{\mu\nu9}&=&\BR_{\mu\nu} +\frac{2\BR_{9[\mu}
	\jmath_{\nu]9}}{\jmath_{99}}\,, \label{eq:TdC9}\\
  C_{\mu\nu\rho}&=&D_{9\mu\nu\rho}
	+\frac{3}{2}\,\Bigl\{\Bigl(\BNS_{9[\mu}\,\BR_{\nu\rho]}
	+\frac{2\BNS_{9[\mu}\,\BR_{|9|\nu}\jmath_{\rho]9}}%
	{\jmath_{99}}\Bigr)-(1\leftrightarrow2)\Bigr\}\,,\\
  B_{\mu\nu}&=& \BNS_{\mu\nu}
	+\frac{2\BNS_{9[\mu}\jmath_{\nu]9}}{\jmath_{99}}\,,\\
  B_{9\mu}&=& \frac{\jmath_{9\mu}}{\jmath_{99}}\,,\\
  A_\mu &=& -\BR_{9\mu} + l \BNS_{9\mu}\,,\\
  A_9 &=& l\,,\\
  \phi&=&\varphi -\frac{1}{2}\ln \jmath_{99}\,.\label{eq:Tdphi}
\end{eqnarray}

\subsection{Metric in the rotated coordinate}
On the other hand, the 9-10 rotated metric is given by
($\RM,\RN=0,1,\cdots,8,y,z$)
\begin{eqnarray}
 \tG_{\RM\RN} &=& \hG_{MN}\,\frac{\p X^M}{\p X^{\RM}}\,
	\frac{\p X^N}{\p X^{\RN}}\nn
  &=& \left(\begin{array}{@{\,}ccc@{\,}}
	\frac{1}{\sqrt{\tG_{zz}}}\,\tg_{\mu\nu}
	+\frac{1}{\tG_{zz}}\tG_{\mu z}\tG_{\nu z}&
	\frac{1}{\sqrt{\tG_{zz}}}\,\tg_{\mu y}
	+\frac{1}{\tG_{zz}}\tG_{\mu z}\tG_{yz}& \tG_{\mu z} \\[10pt]
	\frac{1}{\sqrt{\tG_{zz}}}\,\tg_{y\nu}
	+\frac{1}{\tG_{zz}}\tG_{yz}\tG_{\nu z}&
	\frac{1}{\sqrt{\tG_{zz}}}\,\tg_{yy}
	+\frac{1}{\tG_{zz}}\tG_{yz}\tG_{yz}& \tG_{yz} \\[10pt]
	\tG_{\nu z}&\tG_{yz} & \tG_{zz}\end{array}\right)\,.
\end{eqnarray}
Thus, we have
\begin{eqnarray}
 \tG_{zz}&=&\hq^2\,\hG_{99}+2\hp\hq\,\hG_{910}
		+\hp^2\,\hG_{1010}
  =e^{4\varphi/3}\jmath_{99}^{-2/3}\left\{(\hp+\hq l)^2
	+e^{-2\varphi}\hq^2\right\}\,,\label{eq:2zz}\\
 \tG_{yy}&=&e^{4\varphi/3}\jmath_{99}^{-2/3}\left\{(\hq-\hp l)^2
	+\hp^2e^{-2\varphi}\right\}\,,\label{eq:2yy}\\
 \tG_{yz}&=&e^{4\varphi/3}\jmath_{99}^{-2/3}\left\{
	(\hp l-\hq)(\hq l+\hp)
	+\hp\hq e^{-2\varphi}\right\}\,.\label{eq:2yz}
\end{eqnarray}
Furthermore,
\begin{equation}
 \tG_{\mu y}= \hp\,\hG_{\mu9}-\hq\,\hG_{\mu10}
	 = \frac{1}{\sqrt{\tG_{zz}}}\,\tg_{\mu y}
	+\frac{1}{\tG_{zz}}\tG_{\mu z}\tG_{yz}\,,
\end{equation}
and hence
\begin{eqnarray}
 \tg_{\mu y}&=&\frac{1}{\sqrt{\tG_{zz}}}\left(
	\tG_{\mu y}\tG_{zz}-\tG_{\mu z}\tG_{yz}\right)
  =\frac{\Bpq_{9\mu}}{\jmath_{99}}\,,
\end{eqnarray}
where, as they are given in (\ref{eq:Dpq}) and (\ref{eq:Bpq}),
\begin{equation}
 \Dpq=\sqrt{(\hp+\hq l)^2+e^{-2\varphi}\hq^2}\,,\quad
  \Bpq_{\hmu\hnu}=\Dpq^{-1}(\hp\BNS_{\hmu\hnu}+\hq\BR_{\hmu\hnu})\,.
\end{equation}
Similarly, we have
\begin{eqnarray}
 \tg_{yy}&=&\frac{1}{\Dpq\jmath_{99}}\,,\\
 \tg_{\mu\nu}&=&\Dpq\Bigl(\jmath_{\mu\nu}
	-\frac{\jmath_{9\mu}\jmath_{9\nu}}{\jmath_{99}}
    +\,\frac{\Bpq_{9\mu}\Bpq_{9\nu}}{\jmath_{99}}\Bigr) \,.
\end{eqnarray}
Note that
\begin{equation}
 \sqrt{\frac{\tG_{zz}}{\hG_{1010}}}=\Dpq\,.
\end{equation}

\section{IIA action (\ref{eq:IIA})}\label{S:Cal2a}
First, the $\theta^0$-order part of (\ref{eq:Macr2}) is given by
\begin{equation}
 S_{ddr}\vert_{\theta^0}=2\pi T\int\!d^2 \sigma
  \,{\zl}\Biggl[-\sqrt{-\tg}\,
	-\frac{1}{2}\,\epn^{ij}\tA_{ji}\Biggr]\,.
\end{equation}
Next, we shall calculate the $\theta^2$-order part of
(\ref{eq:Macr2}). Due to
\begin{equation}
 \bar{\theta}_+\Gamma_{\hr}\Gamma^{\hu\hv}\Gamma_{\hs}\theta_-
	=-\bar{\theta}_-\Gamma_{\hs}\Gamma^{\hu\hv}
	\Gamma_{\hr}\theta_+\,,\quad
 \bar{\theta}_+\Gamma_{\hr}\Gamma^{\hu_1\hu_2\hu_3\hu_4}
	\Gamma_{\hs}\theta_-=\bar{\theta}_-\Gamma_{\hs}
	\Gamma^{\hu_1\hu_2\hu_3\hu_4}\Gamma_{\hr}\theta_+\,,
\end{equation}
we have
\begin{eqnarray}
 Q^{(2)}_{\hr\hs}&=&-\frac{i}{4\cdot n!}\,e^{\tphi}\,\bar{\theta}
	\Gamma_{(\hr}\Gamma^{\hu\hv}\Gamma_{\hs)}\Gamma^{10}\theta\,
	\tF_{\hu\hv}\,,\nn
 Q^{(4)}_{\hr\hs}&=&\frac{i}{4\cdot n!}\,e^{\tphi}\,\bar{\theta}
	\Gamma_{(\hr}\Gamma^{\hr_1\cdots\hr_4}\Gamma_{\hs)}\theta\,
	\tF_{\hr_1\cdots\hr_4}\,,\nn
 P^{(2)}_{\hr\hs}&=&\frac{i}{4\cdot n!}\,e^{\tphi}\,\bar{\theta}
	\Gamma_{[\hr}\Gamma^{\hu\hv}\Gamma_{\hs]}\theta\,
	\tF_{\hu\hv}\,,\nn
 P^{(4)}_{\hr\hs}&=&-\frac{i}{4\cdot n!}\,e^{\tphi}\,\bar{\theta}
	\Gamma_{[\hr}\Gamma^{\hr_1\cdots\hr_4}\Gamma_{\hs]}
	\Gamma^{10}\theta\,\tF_{\hr_1\cdots\hr_4}\,.
\end{eqnarray}
Then, the $\theta^2$-order part of (\ref{eq:Macr2}) is calculated as
follows,
\begin{eqnarray}
 S_{ddr}|_{\theta^2}&=&\frac{2\pi T}{2}\int\!d^2\sigma
     \,{\zl}\Biggl[-\sqrt{-\tg}\,
    \tg^{ij}(\tQ_{ij}+2i\bar{\theta}\Gamma_{(i}\p_{j)}\theta)
	-\epn^{ij}(-\tP_{ji}-2i\bar{\theta}\Gamma_{[j}
	\Gamma^{10}\p_{i]}\theta)\Biggr]\nn
  &=&\frac{2\pi T}{2}\int\!d^2\sigma
    \,{\zl}\Biggl[-\sqrt{-\tg}\,\tg^{ij}\Bigl\{
	2i\bar{\theta}\Gamma_i\p_j\theta
	+\frac{i}{2}\,\bar{\theta}\Gamma_i\Gamma_{\hr\hs}
	\theta\tomg_j^{~\hr\hs}\nn
  &&{}\quad\quad-\frac{i}{8}\,e^{\tphi}\bar{\theta}\Gamma_i
	\Gamma^{\hr\hs}\Gamma_j\Gamma^{10}\theta\tF_{\hr\hs}
    -\frac{i}{4\cdot4!}\,e^{\tphi}\bar{\theta}\Gamma_i
	\Gamma^{\hrstu}\Gamma_j\theta\tF_{\hrstu}
    +\frac{i}{4}\,\bar{\theta}
    \Gamma_i\Gamma^{\hr\hs}\Gamma^{10}\theta\tH_{j\hr\hs}\Bigr\}\,\nn
  &&\quad-\epn^{ij}\Bigl\{
      2i\bar{\theta}\Gamma_i\Gamma^{10}\p_j\theta
	+\frac{i}{2}\,\bar{\theta}\Gamma_i\Gamma_{\hr\hs}\Gamma^{10}
	\theta\tomg_j^{~\hr\hs}
      +\frac{i}{8}\,e^{\tphi}\bar{\theta}\Gamma_i
	\Gamma^{\hr\hs}\Gamma_j\theta\tF_{\hr\hs}\nn
  &&{}\qquad+\frac{i}{4\cdot4!}\,e^{\tphi}\bar{\theta}\Gamma_i
	\Gamma^{\hrstu}\Gamma_j\Gamma^{10}\theta\tF_{\hrstu}
	+\frac{i}{4}\,\bar{\theta}
	\Gamma_i\Gamma^{\hr\hs}\theta\tH_{j\hr\hs}\Bigr\}\Biggr]\nn
  &=&-2\pi iT\int\!d^2 \sigma\,{\zl}\bar{\theta}
      (\sqrt{-\tg}\,\tg^{ij}
	-\epn^{ij}\Gamma^{10})\Gamma_i\nabla_j\theta\nn
  &=&-2\pi iT\int\!d^2 \sigma\,{\zl}\sqrt{-\tg}\,\tg^{ij}\bar{\theta}
      \Bigl(1-\frac{1}{2\sqrt{-\tg}}\,\epn^{kl}
	\Gamma_{kl}\Gamma^{10}\Bigr)\Gamma_i\nabla_j\theta\,,
\end{eqnarray}
where
\begin{eqnarray}
  D_i&=&\p_i+\frac{1}{4}\,\p_iX^{\rmu}\Gamma_{\hr\hs}\,
	\tomg_{\rmu}^{~\hr\hs}\,,\\
  \nabla_i&=&D_i-\frac{1}{16}\,e^{\tphi}
	\Gamma^{\hr\hs}\Gamma_j\Gamma^{10}\tF_{\hr\hs}
      -\frac{1}{8\cdot4!}\,e^{\tphi}
	\Gamma^{\hrstu}\Gamma_j\tF_{\hrstu}
	+\frac{1}{8}\,\Gamma^{\hr\hs}\Gamma^{10}H_{j\hr\hs}\,.
\end{eqnarray}

\section{IIB action (\ref{eq:pqacNG})}\label{S:Cal2b}
Note that
\begin{equation}
 \bbtheta_1\Gamma_{\hr}\Gamma^{\hu_1\cdots\hu_n}
	\Gamma_{\hs}\,\btheta_2
  =(-)^{\frac{n+1}{2}}\bbtheta_2\Gamma_{\hs}\Gamma^{\hu_1\cdots\hu_n}
	\Gamma_{\hr}\,\btheta_1\,.\quad(n=1,3,5)
\end{equation}
Then, the $\theta^2$-order part of the action (\ref{eq:pqac2}) is
calculated in the matrix form (cf. (\ref{eq:sp2b})) as follows,
\begin{eqnarray}
 S_{ddr}\big|_{\theta^2}&=&-2\pi iT\!\int\!\!d^2\sigma\,\zl
    \Biggl[\Dpq\bbtheta(\sqrt{-\jmath}
    \jmath^{ij}+\sigma_3\otimes\epn^{ij})\GB_i\DB_j\btheta\nn
  &&{}+\frac{1}{2}\Biggl\{\frac{1}{4}\,\bbtheta
	(\sigma_3\otimes\sqrt{-\jmath}\jmath^{ij}+\epn^{ij})
    \GB_i\GB^{\hrho\hsigma}\btheta\hth_{j\hrho\hsigma}\nn
 &&{}\quad+\frac{e^{\varphi(1)}}{4}\,
      \bbtheta\,\{i\sigma_2\otimes\sqrt{-\jmath}\jmath^{ij}
	+\sigma_1\otimes\epn^{ij}\}\,\GB_i
	\GB^{\hmu_1}\GB_j\btheta\fo_{\hmu_1}\nn
  &&{}\quad-\frac{e^{\varphi(3)}}{4\cdot 3!}\,\bbtheta
    \{\sigma_1\otimes\sqrt{-\jmath}\jmath^{ij}
	+i\sigma_2\otimes\epn^{ij}\}\,\GB_i
    \GB^{\hmu_1\hmu_2\hmu_3}\GB_j\btheta\fth_{\hmu_1\hmu_2\hmu_3}\nn
  &&{}\quad-\frac{e^{\varphi(5)}}{8\cdot 5!}\,\bbtheta
      \{i\sigma_2\otimes\sqrt{-\jmath}\jmath^{ij}
	+\sigma_1\otimes\epn^{ij}\}\,\GB_i\GB^{\hmu_1\cdots\hmu_5}
	\GB_j\btheta\ffv_{\hmu_1\cdots\hmu_5}\nn
 &&{}\quad-\frac{1}{2}\,\epn^{ij}
	\bbtheta\sigma_3\GB_i\GB^{\hmu}\GB_j\btheta
	\p_{\hmu}\Dpq\Biggr\}\Biggr]\nn
 &=&-2\pi iT\!\int\!\!d^2\sigma\,\zl
    \Biggl[\Dpq\bbtheta(\sqrt{-\jmath}
    \jmath^{ij}+\sigma_3\otimes\epn^{ij})\GB_i\DB_j\btheta\nn
  &&{}+\frac{1}{2}\Biggl\{\frac{1}{4}\,\bbtheta
	(\sqrt{-\jmath}\jmath^{ij}+\sigma_3\otimes\epn^{ij})
	(\sigma_3\otimes\GB_i\GB^{\hrho\hsigma})
	\btheta\hth_{j\hrho\hsigma}\nn
 &&{}\quad+\frac{e^{\varphi(1)}}{4}\,
      \bbtheta\,(\sqrt{-\jmath}\jmath^{ij}
	+\sigma_3\otimes\epn^{ij})(i\sigma_2\otimes\GB_i
	\GB^{\hmu_1}\GB_j)\btheta\fo_{\hmu_1}\nn
  &&{}\quad-\frac{e^{\varphi(3)}}{4\cdot 3!}\,\bbtheta
    (\sqrt{-\jmath}\jmath^{ij}+\sigma_3\otimes\epn^{ij})
	\,(\sigma_1\otimes\GB_i\GB^{\hmu_1\hmu_2\hmu_3}
	\GB_j)\btheta\fth_{\hmu_1\hmu_2\hmu_3}\nn
  &&{}\quad-\frac{e^{\varphi(5)}}{8\cdot 5!}\,\bbtheta
      (\sqrt{-\jmath}\jmath^{ij}+\sigma_3\otimes\epn^{ij})
	(i\sigma_2\otimes\GB_i\GB^{\hmu_1\cdots\hmu_5}
	\GB_j)\btheta\ffv_{\hmu_1\cdots\hmu_5}\nn
 &&{}\quad-\frac{1}{2}\,\bbtheta
	(\sqrt{-\jmath}\jmath^{ij}+\sigma_3\otimes\epn^{ij})
	\GB_i\GB^{\hmu}\GB_j\btheta\p_{\hmu}\Dpq\Biggr\}\Biggr]\nn
  &=&-4\pi iT\!\int\!\!d^2\sigma\,\zl\Dpq
    \sqrt{-\jmath}\jmath^{ij}\bbtheta\PB\GB_i\nB_j\btheta\,,
\label{eq:2bpqac0}
\end{eqnarray}
where
\begin{eqnarray}
 \PB&=&\frac{1}{2}\Bigl(1+\frac{1}{2\sqrt{-\jmath}}\,\sigma_3
	\epn^{ij}\GB_{ij}\Gamma^{10}\Bigr),\\
 \DB_i&=&\p_i+\frac{1}{4}\,\p_i\XB^{\hmu}\,
	\omega_{\hmu}^{~\hr\hs}\Gamma_{\hr\hs}\,,\\
 \nB_i&=&\DB_i-\frac{1}{4}\,\p_{\hmu}(\ln\Dpq)\GB^{\hmu}\GB_i
	+\frac{1}{8\Dpq}\,\sigma_3\otimes
	\GB^{\hrho\hsigma}\hth_{i\hrho\hsigma}\nn
 &&{}+\frac{e^{\varphi}}{8}\Bigl(
	i\sigma_2\otimes\Dpq^2\,\GB^{\hmu_1}\fo_{\hmu_1}
      -\sigma_1\otimes\frac{\Dpq}{3!}\,
	\GB^{\hmu_1\hmu_2\hmu_3}\fth_{\hmu_1\hmu_2\hmu_3}\nn
  &&{}\qquad-i\sigma_2\otimes\frac{1}{2\cdot 5!}\,
    \GB^{\hmu_1\cdots\hmu_5}
	\ffv_{\hmu_1\cdots\hmu_5}\Bigr)\,\GB_i\,,
\end{eqnarray}
and use has been made of the relations
\begin{equation}
 \jmath^{ij}\bbtheta\GB_i\GB^{\hmu}\GB_j\btheta=0\,,\quad
 \jmath^{ij}\epn^{kl}\GB_{kl}\GB_i=2\epn^{ij}\GB_i\,.
\end{equation}

\section{\boldmath$\kappa$ parameters (\ref{eq:kpX9})}\label{S:Calkp}
Let $\lambda^i$ and $\gamma^i_{\pm j}$ be
\begin{equation}
 \lambda^i\equiv\frac{\epn^{ij}}{\sqrt{-\tg}}\,
	\Gamma_j\Gamma^{10}\kappa\,,\quad
 \gamma^i_{\pm j}\equiv\frac{1}{2}\Bigl(\delta^i_j\pm
	\frac{\epn^{ik}\tg_{kj}}{\sqrt{-\tg}}\Bigr)\,.
\end{equation}
Note that $\gamma^i_{\pm j}$ satisfies
\begin{equation}
  \gamma^i_{\pm j}\gamma^j_{\pm k}=\gamma^i_{\pm k}\,,
  \quad \gamma^i_{\pm j}\gamma^j_{\mp k}=0\,,
  \quad \gamma^i_{+j}+\gamma^i_{-j}=\delta^i_j\,,
\end{equation}
and also
\begin{equation}
 \epn^{ij}\gamma^k_{\pm i}\gamma^l_{\pm j}=0\,,\quad
  \epn^{ij}\gamma^k_{\pm i}\gamma^l_{\mp j}
	=\frac{\epn^{kl}\pm\sqrt{-\tgamma}\tgamma^{kl}}{2}\,,\quad
 \epn^{ij}\gamma^k_{-i}\gamma^l_{+j}=\epn^{kj}\gamma^l_{+j}
	=\epn^{il}\gamma^k_{-i}\,.\label{eq:Gid}
\end{equation}
Then, we have
\begin{equation}
 \Gamma_i(\gamma^i_{+j}\,\Gamma_-+\gamma^i_{-j}\,\Gamma_+)\lambda^j
	=(1+\Gamma_F)\kappa\,,\label{eq:CalP0}
\end{equation}
or
\begin{equation}
 \Gamma_i\,\gamma^i_{\pm j}\,\lambda^j_{\mp}
	=(1+\Gamma_F)\kappa_{\pm}\,.\label{eq:lamkap}
 \quad(\lambda^i_{\pm}\equiv\Gamma_{\pm}\lambda^i)
\end{equation}
We should note that the projection $\gamma^i_{\pm j}$ can be written,
up to the equations of motion of the auxiliary variables, by
\begin{equation}
 \gamma^i_{\pm j}=\frac{1}{2}\Bigl(\delta^i_j\pm
	\frac{\epn^{ik}\tgamma_{kj}}{\sqrt{-\tgamma}}\Bigr)
	=\frac{1}{2}\Bigl(\delta^i_j\pm
	\frac{\epn^{ik}\jmath_{kj}}{\sqrt{-\jmath}}\Bigr)\,.
\end{equation}

Since $\Xi_i$ in eq.(\ref{eq:d-Y}) can be calculated as
\begin{eqnarray}
 \Xi_i
  &=&-\gamma^j_{+i}\Dpq\Bigl(\p_j\XB^{\hmu}\jmath_{9\hmu}
    +\Bpq_{9\hmu}\,\p_j\XB^{\hmu}\Bigr)\nn
  &&{}+\gamma^j_{-i}\,\Dpq\Bigl(\p_j\XB^{\hmu}\jmath_{9\hmu}
    -\Bpq_{9\hmu}\,\p_j\XB^{\hmu}\Bigr)+O(\theta^2)\,,\label{eq:Xi}
\end{eqnarray}
we have
\begin{equation}
 \Gamma_i=\p_iX^{\rmu}\tse_{\rmu}^{~\hr}\Gamma_{\hr}
	=\Dpq^{1/2}(\gamma^j_{-i}\GB_j-\gamma^j_{+i}\Op\GB_j\Op)\,,
\label{eq:Gi}
\end{equation}
where use has been made of $\Xi_i\simeq\p_iX^y$. And then,
\begin{equation}
 \Gamma_i(\gamma^i_{+j}\Gamma_-+\gamma^i_{-j}\Gamma_+)\lambda^j
 =\Dpq^{1/2}(\gamma^i_{-j}\GB_i\Gamma_+
    -\gamma^i_{+j}\Op\GB_i\Gamma_+\Op)\lambda^j\,.\label{eq:CalP1}
\end{equation}
Due to (\ref{eq:Gi}) $\lambda^i$ is rewritten by
\begin{equation}
 \lambda^i=\frac{\epn^{ij}}{\sqrt{-\tg}}\,\Dpq^{1/2}
	(\gamma^k_{-j}\GB_k-\gamma^k_{+j}\Op\GB_k\Op)
	\Gamma^{10}\kappa\,,
\end{equation}
and hence
\begin{equation}
 \gamma^l_{+i}\lambda^i=\frac{\Dpq^{1/2}\epn^{ij}}{\sqrt{-\tg}}\,
	\gamma^l_{+i}\GB_j\Gamma^{10}\kappa\,,\quad
 \gamma^l_{-i}\lambda^i=-\frac{\Dpq^{1/2}\epn^{ij}}{\sqrt{-\tg}}\,
	\gamma^l_{-i}\Op\GB_j\Op\Gamma^{10}\kappa\,.
\end{equation}

Now that we shall evaluate $\epn^{ij}\Gamma_{ij}$. We have
\begin{equation}
 \epsilon^{ij}\Gamma_{ij}=\Dpq\epn^{ij}\GB_{ij}
    \Bigl(1+\frac{2\epn^{ij}\gamma^k_{-i}\gamma^l_{+j}\jmath_{l9}}%
	{\sqrt{-\jmath}\jmath_{99}}\,
	\{\GB_{k9}+\jmath_{k9}(1+\Gamma_B\Gamma^{10})\}\Bigr)\,,
\label{eq:epnG}
\end{equation}
where
\begin{equation}
 \jmath_{i9}\equiv\p_i\XB^{\hmu}\jmath_{\hmu 9}\,,
	\quad \GB_{i9}\equiv\p_i\XB^{\hmu}\GB_{\hmu 9}\,.
\end{equation}
Up to the equations of motion of the auxiliary variables, we also
have
\begin{equation}
 \tg_{ij}=\Dpq\,\jmath_{ij}\Bigl(1-\jmath^{kl}\,\p_k\XB^{\hmu}
	\p_l\XB^{\hnu}\,\frac{\jmath_{9\hmu}\jmath_{9\hnu}}%
	{\jmath_{99}}\Bigr)+O(\theta^2)\,.\label{eq:tg1}
\end{equation}
Thus, $\Gamma_F$ (\ref{eq:GF}) is written by
(cf. (\ref{eq:epnG}))
\begin{equation}
	\Gamma_F=\Gamma_B\,\calJ\,,
\end{equation}
where
\begin{equation}
 \calJ\equiv\frac{1+\frac{2\epn^{ij}\gamma^k_{-i}
	\gamma^l_{+j}\jmath_{l9}}{\sqrt{-\jmath}\jmath_{99}}\,
	\{\GB_{k9}+\jmath_{k9}(1+\Gamma_B\Gamma^{10})\}}%
	{1-\jmath^{kl}\jmath_{99}^{-1}
	\jmath_{k9}\jmath_{l9}}+O(\theta^2)\,.\label{eq:calJ}
\end{equation}

Now that we shall calculate the projection,
\begin{equation}
 1+\Gamma_F=\frac{1+\Gamma_B}{2}\,(1+\calJ)
	+\frac{1-\Gamma_B}{2}\,(1-\calJ)\,.\label{eq:CndcalA}
\end{equation}
First of all, we have
\begin{equation}
 1-\calJ=\frac{1-\Gamma_B\Gamma^{10}}{2}\,\frac{2\epn^{ij}
	\gamma^k_{-i}\gamma^l_{+j}\jmath_{l9}\,\GB_9\GB_k}%
	{\sqrt{-\jmath}\jmath_{99}(1-\jmath^{kl}\jmath_{99}^{-1}
	\jmath_{k9}\jmath_{l9})}+O(\theta^2)\,,\label{eq:1mJ}
\end{equation}
where use has been made of (\ref{eq:Gid}) and
\begin{equation}
 \gamma^i_{-k}(1+\Gamma_B\Gamma^{10})\GB_i
    =\gamma^i_{+k}(1-\Gamma_B\Gamma^{10})\GB_i=0\,.\label{eq:GB1}
\end{equation}
Thus we have
\begin{equation}
  (1+\Gamma_B\Gamma^{10})(1-\calJ)=0\,,
\end{equation}
and (\ref{eq:CndcalA}) leads to
\begin{equation}
 (1+\Gamma_F)\Gamma_-
    =\frac{1+\Gamma_B}{2}\,(1+\calJ)\Gamma_-\,.\label{eq:kpB1}
\end{equation}
In addition, we have
\begin{equation}
 (1-\Op\Gamma_F\Op)\Gamma_-
    =\{1-\Op\Gamma_B\Gamma^{10}\Op-\Op(1-\calJ)\Op\}\Gamma_-\,.
\end{equation}
Since we have
\begin{eqnarray}
 \Op\Gamma_B\Gamma^{10}\Op
  &=&-\Gamma_B\Gamma^{10}+\frac{2\epn^{ij}\jmath_{j9}}%
	{\sqrt{-\jmath}\sqrt{\jmath_{99}}}\GB_i\Gamma^{10}\Op\,,\\
 \Op(1-\calJ)\Op
  &=&\Bigl(1-\Gamma_B\Gamma^{10}+\frac{2\epn^{mn}\jmath_{n9}}%
	{\sqrt{-\jmath}\sqrt{\jmath_{99}}}\GB_m\Gamma^{10}\Op\Bigr)\,
	\frac{\epn^{ij}\gamma^k_{-i}\gamma^l_{+j}\jmath_{l9}\,
	\Gamma^{10}\GB_k\Op}{\sqrt{-\jmath}\sqrt{\jmath_{99}}
	(1-\jmath^{kl}\jmath_{99}^{-1}\jmath_{k9}\jmath_{l9})}\,,\nn
\end{eqnarray}
we obtain
\begin{equation}
 (1-\Op\Gamma_F\Op)\Gamma_-
  =(1-\Gamma_B)\,\frac{1-\frac{\epn^{mn}\jmath_{n9}\GB_m\Op}%
    {\sqrt{-\jmath}\sqrt{\jmath_{99}}}}{1-\jmath^{kl}\jmath_{99}^{-1}
	\jmath_{k9}\jmath_{l9}}\,\Gamma_-\,,\label{eq:kpB2}
\end{equation}
where use has been made of (\ref{eq:GB1}) and
\begin{eqnarray}
 &&\epn^{mn}\jmath_{n9}\GB_m\Gamma^{10}\Op
	\Bigl(1+\frac{\epn^{ij}\gamma^k_{-i}\gamma^l_{+j}\jmath_{l9}\,
	\Gamma^{10}\GB_k\Op}{\sqrt{-\jmath}\sqrt{\jmath_{99}}
	(1-\jmath^{kl}\jmath_{99}^{-1}\jmath_{k9}\jmath_{l9})}\Bigr)\nn
  &&=\pbp\,\frac{\epn^{mn}\jmath_{n9}\GB_m
	\Gamma^{10}\Op -\frac{\sqrt{-\jmath}\jmath^{nl}\jmath_{n9}
	\jmath_{l9}}{\sqrt{\jmath_{99}}}}%
	{1-\jmath^{kl}\jmath_{99}^{-1}\jmath_{k9}\jmath_{l9}}
      +\pbm\,\frac{\epn^{mn}\jmath_{n9}
	\GB_m\Gamma^{10}\Op}%
	{1-\jmath^{kl}\jmath_{99}^{-1}\jmath_{k9}\jmath_{l9}}\,,
\end{eqnarray}
where
\begin{equation}
 \pbpm\equiv \frac{1\pm\Gamma_B\Gamma^{10}}{2}\,.
	\qquad(\pbpm^2=\pbpm,\quad\pbpm\pbmp=0)
\end{equation}
That is, eqs.(\ref{eq:kpB1}) and (\ref{eq:kpB2}) lead to
eq.(\ref{eq:kpX9}).



\begin{thebibliography}{99}
\bibitem{BST}E.~Bergshoeff, E.~Sezgin and P.~K.~Townsend,
	``Supermembranes And Eleven-Dimensional Supergravity,''
	Phys.\ Lett.\ B {\bf 189}, 75 (1987).

\bibitem{DHIS}M.~J.~Duff, P.~S.~Howe, T.~Inami and K.~S.~Stelle,
	``Superstrings In D = 10 From Supermembranes In D = 11,''
	Phys.\ Lett.\ B {\bf 191}, 70 (1987).

\bibitem{Bus}
  T.~H.~Buscher,
  ``A Symmetry of the String Background Field Equations,''
  Phys.\ Lett.\  B {\bf 194}, 59 (1987).
  ``Path Integral Derivation of Quantum Duality in Nonlinear Sigma
  Models,''
  Phys.\ Lett.\  B {\bf 201}, 466 (1988).

\bibitem{BHO} E.~Bergshoeff, C.~M.~Hull and T.~Ortin,
	``Duality in the type II superstring effective action,''
	Nucl.\ Phys.\ B {\bf 451}, 547 (1995) [arXiv:hep-th/9504081].

\bibitem{MO} P.~Meessen and T.~Ortin,
	``An Sl(2,Z) multiplet of nine-dimensional type II
	supergravity theories,'' Nucl.\ Phys.\ B {\bf 541}, 195 (1999)
	[arXiv:hep-th/9806120].

\bibitem{CLPS}
  M.~Cvetic, H.~Lu, C.~N.~Pope and K.~S.~Stelle,
  ``T-duality in the Green-Schwarz formalism, and the massless/massive
  IIA duality map,''
  Nucl.\ Phys.\  B {\bf 573}, 149 (2000)
  [arXiv:hep-th/9907202].

\bibitem{dWPP}
  B.~de Wit, K.~Peeters and J.~Plefka,
  ``Superspace geometry for supermembrane backgrounds,''
  Nucl.\ Phys.\  B {\bf 532}, 99 (1998)
  [arXiv:hep-th/9803209].

\bibitem{Sch}J.~H.~Schwarz,
	``An SL(2,Z) Multiplet of Type IIB Superstrings,''
	 Phys.\ Lett.\ B{\bf 360},13 (1995) [arXiv:hep-th/9508143];
	``Superstring Dualities,'' Nucl.\ Phys.\ Proc.\ Suppl.
	{\bf 49} 183 (1996) [arXiv:hep-th/9509148];
	``The power of M theory,''
	Phys.\ Lett.\ B {\bf 367}, 97 (1996) [arXiv:hep-th/9510086].

\bibitem{W} E.~Witten,
	``Bound states of strings and p-branes,''
	Nucl.\ Phys.\ B {\bf 460}, 335 (1996) [arXiv:hep-th/9510135].

\bibitem{T}
  P.~K.~Townsend,
  ``Membrane tension and manifest IIB S-duality,''
  Phys.\ Lett.\  B {\bf 409}, 131 (1997) [arXiv:hep-th/9705160].

\bibitem{CT}
  M.~Cederwall and P.~K.~Townsend,
  ``The manifestly Sl(2,Z)-covariant superstring,''
  JHEP {\bf 9709}, 003 (1997) [arXiv:hep-th/9709002].

\bibitem{OUY1}
  H.~Okagawa, S.~Uehara and S.~Yamada,
  ``(p,q)-string in the wrapped supermembrane on 2-torus: A classical
  analysis of the bosonic sector,''
  Phys.\ Lett.\  B {\bf 639}, 101 (2006)
  [arXiv:hep-th/0603203].

\bibitem{McA}
  I.~N.~McArthur, ``Superspace Normal Coordinates,''
  Class.\ Quant.\ Grav.\  {\bf 1}, 233 (1984).

\bibitem{AD}
  J.~J.~Atick and A.~Dhar,
  ``NORMAL COORDINATES, THETA EXPANSION AND STRINGS ON CURVED
  SUPERSPACE,''
  Nucl.\ Phys.\  B {\bf 284}, 131 (1987).

\bibitem{CFBH}
  E.~Cremmer and S.~Ferrara,
  ``Formulation Of Eleven-Dimensional Supergravity In Superspace,''
  Phys.\ Lett.\  B {\bf 91}, 61 (1980);\\
  L.~Brink and P.~S.~Howe,
  ``Eleven-Dimensional Supergravity On The Mass-Shell In Superspace,''
  Phys.\ Lett.\  B {\bf 91}, 384 (1980).

\bibitem{MS}
  L.~Martucci and P.~J.~Silva,
  ``On type II superstrings in bosonic backgrounds and their T-duality
  relation,''
  JHEP {\bf 0304}, 004 (2003)
  [arXiv:hep-th/0303102].
  D.~Marolf, L.~Martucci and P.~J.~Silva,
  ``Fermions, T-duality and effective actions for D-branes in bosonic
  backgrounds,''
  JHEP {\bf 0304}, 051 (2003)
  [arXiv:hep-th/0303209].
  D.~Marolf, L.~Martucci and P.~J.~Silva,
  ``The explicit form of the effective action for F1 and D-branes,''
  Class.\ Quant.\ Grav.\  {\bf 21}, S1385 (2004)
  [arXiv:hep-th/0404197].
  D.~Marolf, L.~Martucci and P.~J.~Silva,
  ``Actions and fermionic symmetries for D-branes in bosonic
  backgrounds,''
  JHEP {\bf 0307}, 019 (2003)
  [arXiv:hep-th/0306066].

\bibitem{Hassan}
  S.~F.~Hassan,
  ``T-duality, space-time spinors and R-R fields in curved backgrounds,''
  Nucl.\ Phys.\  B {\bf 568}, 145 (2000) [arXiv:hep-th/9907152].

\bibitem{BT}
  E.~Bergshoeff and P.~K.~Townsend,
  ``Super D-branes,'' Nucl.\ Phys.\  B {\bf 490}, 145 (1997)
  [arXiv:hep-th/9611173].

\bibitem{OUY2}
  H.~Okagawa, S.~Uehara and S.~Yamada,
  ``(p,q)-string in matrix-regularized membrane and type IIB duality,''
  JHEP {\bf 0710}, 053 (2007)
  [arXiv:0708.3484 [hep-th]].
 
\bibitem{UY4} S.~Uehara and S.~Yamada,
	``From supermembrane to super Yang-Mills theory,''
	Nucl.\ Phys.\ B {\bf 696}, 36 (2004)
	[arXiv:hep-th/0405037].

\bibitem{Ced} M.~Cederwall,
        ``Open and winding membranes, affine matrix theory and matrix
        string theory,''
        JHEP {\bf 0212}, 005 (2002) [arXiv:hep-th/0210152].

\bibitem{UY3} S.~Uehara and S.~Yamada,
        ``Wrapped supermembranes, matrix string theory and an infinite
        dimensional Lie algebra,'' arXiv:hep-th/0402012.

\bibitem{SY}Y.~Sekino and T.~Yoneya,
	``From supermembrane to matrix string,''
	Nucl.\ Phys.\ B {\bf 619}, 22 (2001)
	[arXiv:hep-th/0108176];
	T.~Yoneya,
	``From Wrapped Supermembrane to M(atrix) Theory,''
	arXiv:hep-th/0210243.

\bibitem{UY} S.~Uehara and S.~Yamada,
	``On the strong coupling region in quantum matrix string
	theory,'' JHEP {\bf 0209}, 019 (2002) [arXiv:hep-th/0207209];
	``On the quantum matrix string,'' arXiv:hep-th/0210261.

\end{thebibliography}
\end{document}